%% file: main.tex
\documentclass[letterpaper,twocolumn,10pt]{article}
\usepackage{usenix2019_v3}

\usepackage{graphicx}
\usepackage{tikz}
\usepackage{amsmath}
\usepackage{amssymb}
\usepackage{pifont}
\usepackage{url}
\usepackage{sectsty}
\usepackage{hyperref}
\usepackage{xspace}
\usepackage{mathbbol}
\usepackage{dsfont}
\usepackage{subcaption}
\usepackage{caption}
\usepackage{listings}
\usepackage[framemethod=TikZ]{mdframed}
\usepackage{multirow}
\usepackage{booktabs}
\usepackage{eso-pic}
\subsubsectionfont{\normalfont\itshape}

\usepackage{bbm}
\usepackage{enumitem}
\usepackage{colortbl}
\usepackage{titlesec}
\usepackage{float}

\definecolor{mygreen}{RGB}{53, 124, 60}
\definecolor{greener}{RGB}{0, 102, 0}
\definecolor{myviolet}{RGB}{184, 69, 239}

\newcommand{\no}{\cellcolor{red!90}}
\newcommand{\yes}{\cellcolor{mygreen!90}}
\newcommand{\discussion}{\cellcolor{orange!80}}
\newcommand{\related}{\cellcolor{cyan!90}}

\newcommand{\dno}{\cellcolor{red!10}}

\newcommand{\dunverified}{\cellcolor{blue!6!magenta!4}}

\newcommand{\filledrect}[1]{%
  \begin{tikzpicture}
    \fill[fill=#1] (0,0) rectangle (0.4,0.2);
  \end{tikzpicture}
}

\newcommand{\sectionref}[1]{\hyperref[#1]{\S\ref*{#1}}}
\newcommand{\boxcol}[1]{\fcolorbox{cyan}{cyan!70}{#1}}
\newcommand{\boxsec}[1]{\fcolorbox{orange}{orange!70}{#1}}
\newcommand{\boxnas}[1]{\fcolorbox{teal!80}{teal!50}{#1}}

\newcommand{\imtiaz}[1]{\textbf{\color{purple}Imtiaz: #1}}

\newcommand*\wcircle[1]{\tikz[baseline=(char.base)]{
            \node[shape=circle,draw,inner sep=1pt] (char) { {\small  #1}};}}

\newcommand*\bcircle[1]{\tikz[baseline=(char.base)]{
            \node[shape=circle,fill,inner sep=1pt] (char) {\textcolor{white}{{\small #1}}};}}

\newcommand*\scircle[1]{\tikz[baseline=(char.base)]{
            \node[shape=circle,fill,inner sep=0pt] (char) {\textcolor{white}{{\small #1}}};}}

\newcommand{\cleft}{\ding{115}}
\newcommand{\cright}{\ding{116}}
\newcommand{\cno}{\ding{55}}
\newcommand{\unsafe}[1]{\textcolor{red}{#1}}

\newcommand{\neutral}[1]{\textcolor{blue!60!magenta!90}{#1}}

\newcommand{\unverified}{\ding{79}}

\definecolor{codegreen}{rgb}{0,0.6,0}
\definecolor{codegray}{rgb}{0.5,0.5,0.5}
\definecolor{codepurple}{rgb}{0.58,0,0.82}
\definecolor{backcolour}{rgb}{0.95,0.95,0.92}

\newcommand{\system}{{\fontfamily{lmss}\selectfont CellularLint}{\fontfamily{cmr}\selectfont}}

\newcommand{\sysmodel}{{\fontfamily{lmss}\selectfont EnCell}{\fontfamily{cmr}\selectfont}}

\newcommand{\layer}[1]{{\fontfamily{pcr}\selectfont #1}{\fontfamily{cmr}\selectfont}}

\newcommand{\myvec}[1]{\mathbf{#1}}

\newcommand{\findingsty}[1]{\textbf{\underline{\textit{Finding #1}}}}

\newcounter{fdcounter}
\newcounter{lteic} 
\newcounter{nric} 
\newcounter{paircount} 
\newcounter{paircountlte} 

\newcolumntype{P}[1]{>{\centering\arraybackslash}p{#1}}
\newcommand{\cmark}{\ding{51}}%
\newcommand{\xmark}{\ding{55}}%
\newcommand{\packet}[1]{{\fontfamily{pcr}\selectfont#1}{\fontfamily{cmr}\selectfont}}

\newcommand{\numlabel}{$7$ }
\newcommand{\numue}{$17$ }
\newcommand{\numtrainour}{$150$ }

\newcommand{\numtestour}{$1881$ }
\newcommand{\numtestournr}{$2541$ }

\newcommand{\numissuesclean}{157}
\newcommand{\numissuesemantic}{186}

\newcommand{\numfindings}{11}

\AddToShipoutPictureBG*{%
 \AtPageUpperLeft{%
   \hspace{\paperwidth}%
   \raisebox{-\baselineskip}{%
     \makebox[0pt][r]{\emph{Extended edition. Paper has been accepted at USENIX Security Symposium 2024} \; \; \; \; \; \; \; \; \;\; \; \; \; \; \; \; 
     \; \; \; \; \; \; \; \;\; \; \; \;}
}}}

\begin{document}

\date{}

\title{\Large \bf \system: A Systematic Approach to Identify Inconsistent Behavior in Cellular Network Specifications }

\renewcommand\footnotemark{}
\renewcommand\footnoterule{}

\author{
{\rm Mirza Masfiqur Rahman*\thanks{*Equal contribution. The student author's name is given first.}, Imtiaz Karim*, Elisa Bertino}\\
Purdue University
}

\maketitle

\begin{abstract}
\label{sec:abstract}
In recent years, there has been a growing focus on scrutinizing the security of cellular networks, often attributing security vulnerabilities to issues in the underlying protocol design descriptions. These protocol design specifications, typically extensive documents that are thousands of pages long, can harbor inaccuracies, underspecifications, implicit assumptions, and internal inconsistencies. In light of the evolving landscape, we introduce~\system--a semi-automatic framework for inconsistency detection within the standards of 4G and 5G, capitalizing on a suite of natural language processing techniques. Our proposed method uses a revamped few-shot learning mechanism on domain-adapted large language models. Pre-trained on a vast corpus of cellular network protocols, this method enables \system{} to simultaneously detect inconsistencies at various levels of semantics and practical use cases. In doing so, \system{} significantly advances the automated analysis of protocol specifications in a scalable fashion. In our investigation, we focused on the Non-Access Stratum (NAS) and the security specifications of 4G and 5G networks, ultimately uncovering \numissuesclean{} inconsistencies with $82.67\%$ accuracy. After verification of these inconsistencies on 3 open-source implementations and \numue commercial devices, we confirm that they indeed have a substantial impact on design decisions, potentially leading to concerns related to privacy, integrity, availability, and interoperability. 

\end{abstract}

\input{sections/introduction}
\input{sections/background}

\input{sections/overview}

\input{sections/detailed_design}

\input{sections/implementation}

\input{sections/evaluation}

\input{sections/related_works}

\input{sections/discussion}

\input{sections/conclusion}

\section*{Acknowledgments}
We thank the shepherd and the reviewers for their suggestions and insightful discussion.
The work reported in this paper has
been supported by NSF under grant 2112471.





\bibliographystyle{plain}
{\footnotesize
\bibliography{references}}

\appendix

\input{sections/appendix}
\end{document}

%% file: sections/introduction.tex
\section{Introduction}\label{sec:introduction}















Cellular networks have emerged as the primary means of communication in the modern world, with 4G and 5G being the latest commercially available versions.
Ranging from mobile communications to IoT devices to vehicular communications, cellular networks are a critical 
infrastructure for the digital world. With more than 1.4 billion subscribers, 5G is expected to exceed its much larger predecessor, 4G, and is forecast to gain more than 13 billion subscribers globally by the end of 2026~\cite{gsacom,erics}. These networks embody large infrastructure and support different features, including backward compatibility, interoperability, and heterogeneity. 

Like any large-scale, complex, layered system, cellular network designs are governed by protocol documentation. 3GPP, the 3rd Generation Partnership Project--a consortium of vendors, is the supervising body that handles the design and distribution of these protocol documents~\cite{3GPP}. These documents are quite lengthy, developed by many stakeholders, and improved through releases and change requests over the years. Incorporating these changes and releases subsequently introduces diverging descriptions. Recently, few studies have shown, albeit through manual analysis during implementation testing, that protocol documents have  
inconsistent descriptions accounting for differing design possibilities and undefined behaviors in real-world devices, often with severe consequences~\cite{277258, 10.1145/3460120.3485388}. Fig.~\ref{lt:1.1} shows an example scenario for diverging sub-state transitions on the same condition, found in the 4G Non-Access-Stratum (NAS) specification. Furthermore, in most cases, one of the inconsistent or conflicting behaviors is less secure than the other. For example, if one design suggests clearing some network identifier(s) after a connection has been lost while the other design suggests keeping it without specifying a clear motivation, the latter can make the user vulnerable to privacy violation attacks.
Therefore, these diverging specifications defined in the standards can have severe security and privacy consequences.

\mdfsetup{%
   middlelinecolor=blue!50,
   middlelinewidth=2pt,
   backgroundcolor=blue!20,
   roundcorner=10pt}

\begin{figure}[t]{
    \centering
    \noindent\begin{mdframed}[font=\scriptsize]
    $T_1$: Whenever an ATTACH REJECT message with the EMM cause \textbf{\#14 "EPS
    services not allowed in this PLMN"} is received by the UE $\cdots$ Additionally the attach attempt counter shall be reset when the UE is in substate \textbf{\textcolor{purple!90}{EMMDEREGISTERED.ATTEMPTING-TO-ATTACH}}.
\\
    \noindent $T_2$: \textbf{\#14 (EPS services not allowed in this PLMN}); The UE shall set the EPS update status to EU3 ROAMING NOT ALLOWED $\cdots$  the UE
    shall reset the attach attempt counter and enter the state \textbf{\textcolor{blue}{EMM-DEREGISTERED.PLMN-SEARCH}}.
    \end{mdframed}
    \vspace{-0.5cm}
    \caption{An example inconsistency identified by \system{}-two different sub-state transitions for same precondition. $T_1$ is from section 5.5.1.1 and $T_2$ is from 5.5.1.3.5 of TS 24.301.}
    \label{lt:1.1}}
\end{figure}


\noindent\textbf{Prior Research. } Although prior research works~\cite{10.1145/3319535.3354263,Hussain2018LTEInspectorAS,10.1145/3460120.3485388,277258,291203,9519388,279972,10.1145/3317549.3324927,10.1145/2810103.2813718,10.1145/2810103.2813618} have proposed approaches for detecting security and implementation flaws in cellular network protocols, they have at least one of the following limitations: (A) 
They are completely manual~\cite{10.1145/3460120.3485388,10.1145/3317549.3324927,10.1145/2810103.2813718,10.1145/2810103.2813618} and inherently have limited scalability. 
(B) 
They employ formal verification~\cite{10.1145/3319535.3354263,Hussain2018LTEInspectorAS}, foundationally relying on the quality of the properties, and do not primarily deal with inconsistent descriptions-- subsequently choosing the most secure option when an inconsistent or confusing situation is encountered. (C) They focus on implementation testing~\cite{8835363,277258} and only report a few inconsistent descriptions on the way without proposing an approach to uncover them. (D) They use differential testing for noncompliance checking~\cite{10.1145/3460120.3485388} without focusing on conflicting behavior detection in the standards. (E) They either analyze change requests or specifications through NLP techniques to understand security hazards~\cite{9519388,279972}  without directly considering differing description analysis in specifications.  

\noindent\textbf{Problem. } With the current state-of-the-art approaches having above-mentioned limitations, in this paper, we aim to take the first steps to answer the following research question- \emph{Is it possible to develop a systematic, generalizable, and scalable specification analysis framework that can identify inconsistent descriptions from 4G and 5G specifications and associate them with differential design choices?}

\noindent \textbf{A promising direction. } 
Following the major advances in natural language processing (NLP)~\cite{vaswani2017attention,brown2020language}, recent works have demonstrated the automated generation of Finite State Machine from both Request For Comments (RFC) documents and cellular network specifications~\cite{9833673,ishtiaq2023hermes}. Such results indicate 
that a data-driven approach can be taken to discover inconsistencies in specifications as a root cause of design flaws and vulnerabilities as well.

\noindent \textbf{Challenges. } 
Detecting inconsistent behaviors systematically from the large and complex cellular network specifications using NLP requires tackling some very important research challenges. The first and foremost challenge is to quantize the specification into \emph{segments}, which we can use for inconsistency detection. The specifications explain an event in a very large context; in some cases, an event is described in 5-6 pages. With such a large segment, any learning algorithm would fail to capture the finer details and attributes of the segment. Therefore, we need to devise a mechanism to create segments of reasonable size from the specifications. Second, the specifications are so large that in case all the segments are pairwise compared, the search space becomes intractable. For this, we need to devise an approach that reduces the search space substantially. Last, and most importantly, to the best of our knowledge, there are no datasets that could be utilized to detect inconsistent statements in cellular networks. In case there is a large number of inconsistencies that can be manually found, this trivially solves the problem and does not require a systematic framework. Fortunately or rather unfortunately, there are very few manually found conflicts in the specifications that can be used for closed-form solutions that we can use to train. This lack of ground truth creates a major hurdle for any learning-based solution.


\noindent\textbf{Our Approach. }In this paper, we introduce \system{}--the first framework for uncovering inconsistencies from the upper layer of 4G and 5G specifications. For our method, we employ few-shot domain adaptation of language models. 

To obtain a scalable solution, we initially cast the documents in sizeable segments based on domain-specific knowledge. These quantized segments preserve the context and event, which are essential before understanding inconsistencies. Second, based on this quantized pool of intra-document and inter-document test cases/segments, we produce \emph{Pairs of Segments} (\textbf{PoS}). \system{} compresses the search space of PoS into a manageable and informative dataset based on interpretable similarity measures. 
This guides our method to look into only relevant content--making the process highly scalable.

To address the lack of labeled examples, instead of heavily relying on expert annotations, \system{} incorporates domain-aware annotations and ensemble decision mechanisms and takes only a fraction of labeled examples. We design seven domain-adapted, consistency-wise meaningful labels and map them to general-purpose natural language inference (NLI) annotations. \system{} also utilizes human-in-the-loop active learning on multiple language models to finalize inconsistent descriptions in varying granularities.

To 
address the lack of supervision, we divide the learning objective into multiple phases- (1) initially converting the learning problem to a generalized setting of NLI 
and then (2) using multiple phases of active learning on a domain-specific dataset.

\noindent\textbf{Findings. }\system{} discovered \numissuesclean{} conflicting descriptions from specifications. To evaluate the effectiveness of the discoveries, we investigate them on 3 open-source implementations and test them on \numue commercial UEs. 

We uncover that implementations indeed take different design choices based on conflicting or inconsistent standards. Furthermore, in some cases, we have found that implementations choose the less secure option, 
and in some cases, tried to implement both options. For ease of discussion, we characterize a subset of our findings into four different categories--impacting security enforcement, privacy, interoperability, and causing denial-of-service. In these categories, we discuss a total of \numfindings{} issues that can have a severe impact on the privacy, integrity, availability, and interoperability of cellular network infrastructure. 


\noindent\textbf{Open-source. } We will completely open-source \system{} and all the detected inconsistencies to foster research in inconsistency detection in other important protocol specifications~\cite{cellularlint}~\footnote{\url{https://cellularlint.github.io/}}.

\noindent\textbf{Contributions.}To summarize, we make the following contributions:
\vspace{-0.25cm}
\begin{itemize}[leftmargin=*]
\setlength\itemsep{-0.5em}
    \item[$\blacksquare$] \textit{\textbf{Framework.} } We design and implement \system{}. To the best of our knowledge, this is the first systematic approach towards inconsistency detection for both 4G and 5G specifications.
    \item[$\blacksquare$] \textit{\textbf{Few-shot learner.}} 
    We demonstrate a novel few-shot approach with active multiphase learning for cellular network-specific pre-trained language models to subsidize insufficient labeled data effectively.
    \item[$\blacksquare$] \textit{\textbf{New findings.}} We found \numissuesclean{} inconsistent descriptions and investigated the results on 3 open-source and \numue commercial device implementations.
\end{itemize}

%% file: sections/background.tex
\section{Background}\label{sec:background}
In this section, we introduce various preliminaries ranging from the 4G Long Term Evaluation (LTE) and 5G New Radio (NR) architectures to various NLP methods that are relevant to our work.

\subsection{Cellular Network Preliminary}
For cellular network preliminaries, here we discuss the most common components that are relevant to this paper. 

\subsubsection{Cellular Network Architecture}

The network is composed of three main components: the user equipment (UE), the radio access network (RAN), and the core network. The architectural difference between 4G and 5G mostly lies on the RAN and Core Network side. 

\noindent\textbf{User Equipment (UE).} The UE, a terminal device on the user end, is equipped with a Universal Subscriber Identity  Module (USIM). The USIM contains the user identifier, master secret key, and session key for connectivity. These are essential for mutual Authentication and Key Agreement (AKA) between the user and the network. The most common UE are cell phones, tablet computers, and IoT devices with cellular connectivity.

\noindent\textbf{eNodeB.} eNodeBs or eNBs are base stations or, most commonly, mobile network towers in the 4G architecture. They stand as a middle entity in the connection establishment and maintenance for which 
the Radio Resource Control (RRC) protocol is implemented. eNodeB is replaced by gNodeB in the 5G network. eNodeB is responsible for transmitting and receiving signals to and from UEs.

\noindent \textbf{gNodeB.} The gNodeB or gNB--also called the Next-Generation NodeB--is used in 5G network. It is a significant upgrade from eNodeB. gNodeB supports Massive MIMO technology, has higher throughput, lower latency, and enables advanced network slicing. 

\noindent\textbf{4G-Core}
The core network in 4G LTE is called Evolved Packet Core (EPC). Two functional components--Mobility Management Entity (MME) and Home Subscriber Server (HSS) are relevant to our paper; hence we briefly discuss them in what follows.

\begin{itemize}[leftmargin=*]
\setlength\itemsep{-0.2em}
    \item \textbf{Mobility Management Entity (MME).} The MME is responsible for handling signals between the UE and the core, as well as the eNodeB and the core. It is responsible for UE authentication, detach procedure, tracking area updates, etc.
    \item \textbf{Home Subscriber Server.} HSS works as a database to store subscriber information (IMEI and IMSI). It also provides relevant information for calls and IP sessions.
\end{itemize}

\noindent \textbf{5G-Core.} The 5G Core comprises many functional components such as the Access and Mobility Management Function (AMF), Session Management Function (SMF), User Plane Function (UPF), Authentication Server Function (AUSF), and so on. 5G leverages a large number of antennas on the core side to enhance signal quality. 

\subsection{ML Preliminaries}
\system{} detects inconsistencies from a natural language perspective, meaning that it focuses on the lexical and contextual analysis of the salient features of the protocol text documents to reveal inconsistencies from protocol specifications. 
Here, we discuss the NLP preliminaries required for~\system. 

\subsubsection{Few Shot Learning}
Few-Shot Learning (FSL)~\cite{brown2020language, gao-etal-2021-making} is a meta-learning approach that utilizes a pre-trained model to generalize over new categories of data using only a few labeled examples per class.
For $N$-way-$k$-shot learning in NLP, given $k$ labeled examples of each class from $N$ classes for a new NLP task where $k$ is generally very low, the pre-trained model has to learn efficiently to solve the new task. 

\subsubsection{Active Learning}
Active learning~\cite{monarch2021human,dor2020active}, especially human-in-the-loop active learning, is a machine learning paradigm that focuses on improving learning performance by minimizing the need to label large amounts of data. Often, domain-specific supervised learning is hindered by the insufficiency of ground truth data. Guiding the model by associating expert annotation with a subset of important data points can largely boost the performance of supervised training. 

\subsubsection{NLP Methods}

\noindent \textbf{Textual Entailment (TE). }Textual entailment, or Natural Language Inference (NLI), is a binary relation where two text sequences can either form an agreement, a contradiction or be completely irrelevant to each other~\cite{im2017distance,snli:emnlp2015}. From a NLP perspective, given two text sequences, where the first is a hypothesis and the second is a premise, the model has to find out whether the first text is in agreement with the second (or vice versa) or is in contradiction or has no logical relation.

\noindent \textbf{Language Modeling.} Recent years have seen revolutionary improvements in machine learning on textual data, much of which is due the use of transformers~\cite{vaswani2017attention,wolf-etal-2020-transformers,DBLP:journals/corr/abs-1810-04805,DBLP:journals/corr/abs-1907-11692,yang2019xlnet}. 
Transformers are widely used for sequence-to-sequence modeling, semantic parsing, machine translation, question-answering, and 
most recently for large language models. 
Transformers rely on the attention mechanism, which excels at learning complex linguistic patterns effectively through the understanding of each word's importance and its ability to identify its surrounding words, thereby characterizing the context.


%% file: sections/overview.tex
\section{Overview}\label{sec:overview}

In this section, we discuss the problem analysis scope and state the problem formally. Next, we discuss the challenges and corresponding approaches. 

\subsection{Scope of Analysis}
The cellular network protocols (4G and 5G) comprise thousands of specifications--from the low layers to the link layer to the upper layers (Layer 3) of the protocols. Among all these specifications and related procedures, we focus on the NAS layer procedures. NAS layer procedures contain the mobility management and session management components, which in turn manage the most critical control plane procedures such as connection setup, initial authentication, mobility, hand-off, and service notifications. Previous works have shown several vulnerabilities in these components, resulting in severe security and privacy issues such as authentication-bypass~\cite{8835363}, 
location exposure~\cite{Hussain2019PrivacyAT}, impersonation~\cite{rupprecht-20-imp4gt}, downgrading~\cite{karakoc2023never}, to name a few. 
Keeping these in mind, in this work we focus on the NAS layer procedure. More specifically, we focus on four documents in two categories- the NAS technical specification TS (24.301, v17.6.0)~\cite{nas4g} and the Security Architecture and Procedures specification (TS 33.401, v17.1.0)~\cite{sec4g} for 4G, and the NAS specification (TS 24.501, v17.7.1)~\cite{nas5g} and the Security Architecture and Procedures specification (TS 33.501, v17.5.0)~\cite{sec5g} for 5G. All of them are available from the 3GPP archive. Release 17 is the latest complete version to have both NAS and its security specifications.  Furthermore, having the same release ensures consistency and removes the possibility of unexpected inconsistency prediction (false positives) introduced by version mismatch. In fact, finding cross-version inconsistency is a different task that we leave as future work.


\subsection{Problem Formulation}


Here we define the problem formally: given a set of corpora $\mathcal{G} = \{ \mathcal{C}_{Nas}^{4g}, \mathcal{C}_{Sec}^{4g}, \mathcal{C}_{Nas}^{5g}, \mathcal{C}_{Sec}^{5g}\}$ (where $\mathcal{C}_{Nas}^{4g}$ is the corpus produced from 4G NAS specification, $\mathcal{C}_{Sec}^{4g}$ is the corpus produced from 4G security architecture specification, similar notations are applicable for 5G), we need to quantize each corpus in text segments, that is, $\mathcal{C}_i=\{t_1,t_2,\cdots,t_n\}_i$ where $t_{a}^i$ describes $a$-th meaningful event/sub-event/directive from corpus $\mathcal{C}_i$.  Now our first objective is to construct intermediate sets, $\mathcal{S}_1=(\mathcal{C}_{Nas}^{4g}\times\mathcal{C}_{Nas}^{4g})\cup (\mathcal{C}_{Nas}^{4g} \times \mathcal{C}_{Sec}^{4g}) \cup (\mathcal{C}_{Sec}^{4g}\times \mathcal{C}_{Sec}^{4g}$) and $\mathcal{S}_2= (\mathcal{C}_{Nas}^{5g}\times\mathcal{C}_{Nas}^{5g}) \cup (\mathcal{C}_{Nas}^{5g} \times \mathcal{C}_{Sec}^{5g}) \cup (\mathcal{C}_{Sec}^{5g}\times \mathcal{C}_{Sec}^{5g}$). Next, from each intermediate set $\mathcal{S}_m\in\{\mathcal{S}_1,\mathcal{S}_2\}$ containing tuples or pairs $(t_{a}^i,t_{b}^j)$ of text Segments (PoS), our objective is to devise a model $\mathcal{M}$ which can determine 
all pairs that conflict between each other, that is, $\mathcal{M}(t_{a}^i, t_{b}^j) = k$. Here, $k$ takes the value \textit{consistent} or \textit{inconsistent}.


\subsection{Challenges \& Solution Outline}
We now discuss the main challenges and corresponding approaches that have driven the \system{}'s high-level design choices. 
We then present the key steps in our overall solution.

\noindent \textbf{(C1) Segment quantization.} In numerous scenarios, the specifications explain an event in a vast context. For example, in the 4G NAS specification, Section 5.5.1.2.4 describes "Attach accepted by the network", which is a six-page long description concerning one major event where the network accepts an attach request from the UE. If we consider this whole event as one complete and discrete, meaningful \textit{\textbf{segment}}, 
any learning algorithm will fail to capture the intricate details and attributes, 
causing differential and anomalous behaviors. Moreover, in the same section, there is a precondition \textit{"If the UE indicates support for EMM-REGISTERED without PDN connection"}, which shall trigger an event on the MME side. A similar precondition is found for \packet{tracking\_are\_update\_request} message further away in the document. If the whole section is considered as a potential segment, one would not be able to match such sub-event triggers and may miss numerous possible interesting scenarios. 
An alternative choice can be selecting each sentence as an atomic unit governing a segment. However, this leads to a serious problem because, often, a single sentence from the specification is not sufficient to infer all the entities, states, or messages involved in a process. Also, due to the semi-structured nature of specifications, many sentences are repeated in different parts of the specifications. Thus, considering each sentence a meaningful segment would produce too many repeating units in the dataset, and if needed, it would be impossible to map back uniquely to the specification. Thus, this trivial solution is not applicable to our case.

\looseness  = - 1
\noindent \textbf{(A1) Sub-event driven segment quantization.} As we have already established that considering a whole event as a segment likely affects the effectiveness and purpose of detected inconsistencies, we quantize
such larger descriptions into smaller segments. However, in doing so, we preserve the completeness of the description of sub-events. We combine domain-specific understanding with standard NLP techniques to pinpoint sub-events. Specifically, we consider subsections and complete paragraphs as atomic units while constraining the sequence length and entity based on Parts-Of-Speech filtering. Details about the approach for segment quantization are given in~\sectionref{subsec:dp}.

\noindent \textbf{(C2) Intractable search space.} The protocol specifications are large documents defining and explaining all entities, events, state transitions, message structures, and so on in finer detail. Moreover, when we execute segment quantization to address \textbf{C1}, 
the resulting quantized dataset is even larger. 
Thus, traversing and matching texts 
directly using the entire dataset of quantized segments makes the finalized search space intractable with millions of potential segment pairs. For example, after the segment quantization, the brute-force traverse and match solution gives us more than \textit{\textbf{12 million}} segment pairs for 5G-NAS specification alone. 
On top of that, to find inconsistencies, one may have to consider multiple related specifications rather than generating segments from just one. For instance, in our case, for 5G we consider the 5G NAS technical specifications and the security architecture and procedures specification. This, in turn, makes the search space tremendously larger and thereby causes a state-space explosion. 

\noindent \textbf{(A2) Finite segment filtration.} To address \textbf{C2}, we reduce the segment sequence pair space by filtering the initial dataset based on similarity measures. 
The intuition behind this idea is that two completely different segments should not be considered inconsistent as they talk about totally different aspects of the protocol. Contrary to this, an inconsistent PoS would have at least the same pre-condition and hence would have a better similarity score.
To achieve such a quantitative measure of texts, we first consider the vector embedding of texts. However, we observe that choosing an arbitrary embedding can severely affect the search space 
shrinkage and effectiveness of similarity scoring. We have thus carried out an extensive evaluation to determine the most reliable text vectorization technique for our problem. The details of the evaluation are discussed in~\sectionref{subsec:embed}.
Thus we reduce the number of segments from millions to a few thousand--reducing the problem space to a tractable size.


\noindent \textbf{(C3) Absence of ground truth.} 
To the best of our knowledge, there are no datasets that could be utilized for NLP-based machine learning to detect conflicting statements from cellular protocols.
Even if a processed, readily usable dataset existed, no closed-form solution would be able to trivially map text segments from multiple documents to specific inconsistencies. This is a novel problem--solving which is a key contribution of our work. 
Moreover, although state-of-the-art language models are very effective in learning contexts, they require a massive amount of labeled data due to the larger parameter landscape. Acquiring such a domain-specific dataset is often cumbersome and for this problem setup, even harder. Thus we have to design a technique that can work with only a handful of labeled examples.

\noindent \textbf{(A3) Ground truth generation.} We employ multi-phase supervision through human-in-the-loop active learning to subdue the insufficiency of ground truth. Our learning method neither requires a large amount of annotation nor suffers from too much uncertainty introduced by insufficient learning. In short, we first establish the learning landscape using a general-purpose dataset with some supervised data. Note that this general-purpose dataset is not from the cellular network domain but rather from image-crowdsourced caption writing and is publicly available.  
This intermediary model is utilized in consensus with human involvement alongside minimal synthesized data repeatedly to prepare the finalized model. The whole process never requires the complete ground truth of our problem. In fact, having a complete ground truth from the beginning is fundamentally contradictory here, as it would trivially solve the problem we have at hand. Indeed, if there were already a large number of inconsistent descriptions in specifications, then those could be directly reported to improve the protocol and would not require any additional analysis. Likewise, the lack of ground truth makes the problem both challenging and rewarding to some extent.


\noindent \subsection{Approach Skeleton}
Based on our discussion of challenges and approaches to the solution, we divide the inconsistency detection problem into multiple steps: (i) We employ domain-specific preprocessing alongside standard NLP techniques to quantize each $\mathcal{C}_i \in \mathcal{G}$. (ii) To produce each $\mathcal{S}_m$, we utilize an syntactic and contextual equivalence metric $\psi$ to filter out irrelevant 
PoS. (iii) To address the absence of labeled examples, we use $\mathcal{M}'$ instead of $\mathcal{M}$ which first learns from a general dataset $\mathcal{P}$ to solve textual entailment task. In this task, for any two text sequences $t_x,t_y \in \mathcal{P}$ , $\mathcal{M}'$ learns if $t_y$ entails $t_x$ or not. Formally, $\mathcal{M}'(t_x, t_y) = k'$ where $k'$ takes the value of either entailment or contradiction or irrelevance (often called neutral). Subsequently we use $\mathcal{M}'$ with few domain-specific labeled data to generate our specialized model $\mathcal{M}$ with the learning objective of $\mathcal{M}(t^i_a,t^j_b) = k'$, where $t^i_a,t^j_b \in \mathcal{S}_{m}$.

%% file: sections/detailed_design.tex
\section{Detailed Design}\label{sec:detailed design}
\begin{figure*}[ht]
    \centering
    \includegraphics[width= 1\textwidth, trim = 0cm 24cm 0cm 0cm, clip]{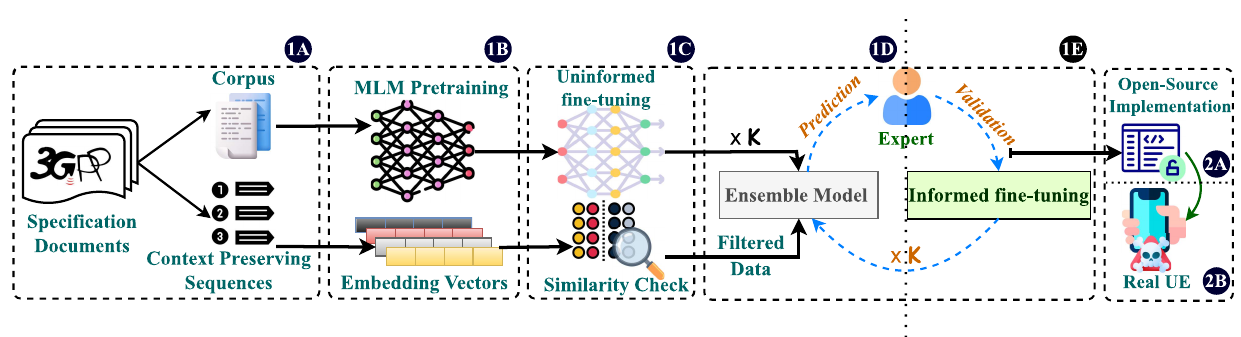}
    \caption{Architecture of \system{}. The two-arrow process shown in \protect \scircle{1A} \protect, \protect \scircle{1B} \protect, and \protect \scircle{1C} \protect represents independent and parallel processing.} 
    
    \label{fig:main-arch}
    \vspace{-0.5cm}
\end{figure*}

In this section, we discuss \system~in detail. Fig.~\ref{fig:main-arch} shows the main components of our framework.
\subsection{Architecture}
\label{subsec:archi}

\noindent 
The overall framework is organized into two parts: \bcircle{1} the Learner, and \bcircle{2} the Dispatcher. Each of them has multiple submodules shown in Figure~\ref{fig:main-arch}. 

\noindent \underline{\textbf{Learner.}} The learner consists of five sub-modules. \\
\noindent\scircle{1A}~The \textbf{\textit{first}} is preprocessing, which performs standard NLP preprocessing as well as domain-specific and task-specific preprocessing of all cellular network protocol specifications from the 3GPP archive. This generates a large corpus for our pre-training step. Here, a second-order preprocessing is also executed on the documents of our problem scope (NAS and Security of 4G and 5G), which produces the sub-event-oriented, context-preserving text segments. We call them context-preserving as when extracting them, our method makes sure that they do not describe multiple sections and the segments are not trimmed off halfway through an event description. For example, if a section is short and within our token limit (such as 5.3.10 in TS 24.301), we keep it as one segment. Alternatively, if a section contains multiple paragraphs (such as 5.5.3.2.3 from TS 24.301), each describing an independent event (they might be related when taken in a very large context), we consider each paragraph as a different segment.  
Details of the module can be found in~\sectionref{subsec:dp}.

\noindent \scircle{1B}~The \textbf{\textit{second}} sub-module 
is the pre-trainer.  In this sub-module, we utilize the large corpus generated from the 
preprocessing module. This ensures that the model understands the domain with specific vocabularies and semantics. The detailed experimental setup is discussed in \sectionref{subsubsec:pt}.
In parallel to the training, we create embedding vectors for each segment. The embedding vectors are utilized for pairing and filtration of PoSs in the next sub-module.

\noindent\scircle{1C}~In the \textbf{\textit{third}} sub-module, we execute the uninformed fine-tuning. This is an important step for weak supervision and our first concrete endeavor toward task-specific learning. In short, we first fine-tune our candidate transformer models separately on the well-known Stanford Natural Language Inference (SNLI) corpus~\cite{snli:emnlp2015}. We call it uninformed supervision because the SNLI corpus contains examples that are not completely in accordance with our definition of consistency from a protocol perspective. For example, "\textit{A group of people are ice skating in a big city.}" and "\textit{The people are outside skating.}" are labeled as entailments in that corpus. Apparently, the second sequence is a vague representation of the first, i.e., a large and significant part of the first sequence is not expressed through the second sequence. Moreover, our main objective is to capture the protocol inconsistencies based on complex text patterns and intricate details of various cellular events for which fine-tuning on SNLI corpus is not solely sufficient.
Thus, training on the SNLI corpus only gives us a loose estimate of our objective model. Nonetheless, this uninformed supervision is the first stepping stone in our analysis. In parallel, we complete the segment \textit{pairing and filtration} based on the embedding vectors generated in the previous step. In short, we create all possible pairs of the segment vectors and filter them based on a similarity measure. Details of the method can be found in \sectionref{subsec:pf}

\noindent\scircle{1D}~At the \textbf{\textit{fourth}} sub-module, we combine the predictions of $k$ models on our dataset. Note that these $k$ models are now fine-tuned on SNLI corpus from the previous step. It is known that different models have the ability to capture different semantics in varying capabilities. Therefore, combining their predictions through majority voting allows us to boost the prediction confidence. However, these models have not learned what consistency means from the perspective of cellular networks. Therefore, these predictions are not completely reliable as final output.

\noindent\scircle{1E}~The most important sub-module for the learner is the multi-phase informed fine-tuning, which is the \textbf{\textit{fifth}} and most crucial step. Here, we first take the ensemble prediction from the previous sub-module. Next, we sample a small subset of data from the predicted set, then validate and rectify the labels through domain-expert human annotators. This approach ensures ground truth addition while keeping the human involvement minimal. Also, rectifying annotations instead of annotating from scratch reduces the human effort even more. The models are again trained on the reconditioned examples in a supervised manner and used subsequently for predictions, again on the whole dataset. In such a manner, we complete one phase of training. This method is followed $k$ times (each defining a phase) before finalizing the prediction.  Details on the experimental choices of this step can be found in \sectionref{subsubsec:ft}.

\noindent \underline{\textbf{Dispatcher.}} We take the high-confidence predictions as our results after multi-phase informed training and manually analyze them. To further show the impact of the inconsistent behaviors of the specifications, we use the dispatcher. 
The dispatcher has two sub-modules.

\noindent\scircle{2A} In the \textbf{\textit{first}} one, we map the discovered inconsistent descriptions to the open-source implementations-srsRAN, open5GS, and OpenAirInterface~\cite{srsRAN,open5GS,openAirInterface}. Note that many boundary cases and uncommon events are not available (or available in non-granularity) in the open-source implementations. Even so, the predicted inconsistent sets are checked against these multiple sources to determine their design choices and security implications. Consequently, we create a subset of inconsistencies for the next sub-module.

\noindent\scircle{2B} In the \textbf{\textit{second}} sub-module of the dispatcher, we consider each of the inconsistencies gathered from the previous step to determine whether they cause issues in real-world devices. Details of the setup can be found on our website~\cite{cellularlint}.



\subsection{Dataset Preparation}\label{subsec:dp}

It has been shown that pre-training Large Language Models (LLM) can help understand domain-specific terminologies better than LLMs with no domain-specific pre-training (i.e., only pre-trained on general datasets such as Wikipedia corpus)~\cite{gu2021domain,chalkidis-etal-2020-legal,rasmy2021med,brown2020language}. 
We first process the raw specifications from the 3GPP archive and pre-train language models on it to leverage the domain-specific learning in an enhanced capability. Our textual entailment task is a harness over the aforementioned pre-trained model. The details of the dataset preparation are discussed in what follows.


We remove the tables, figures, cross-document references, code segments, and additional ill-formed texts from the specifications to create the pre-training corpora. For the downstream fine-tuning, since we need semantically meaningful segments of texts to compare, we first
extract section-wise texts. The sections usually comprise of many sub-descriptions based on cause values. Otherwise, each paragraph in a section is usually self-contained. Also, to conform to the sequence length limit of our candidate transformer models, we sometimes do finer fragmentation. 
In this case, sections 4 to 8 are considered for NAS in both 4G and 5G as the rest of the sections mostly contain definitions, abbreviations, glossary, scope, etc. For security, we consider section 4 to the annex for both 4G and 5G. 

\subsection{Pairing \& Filtration}\label{subsec:pf}
As discussed earlier, comparing all segments with each other for inconsistency will result in a massive search space, quantitatively, $N\choose 2$ datapoints where $N$ is the total number of extracted segments. To overcome this, first, we use the Term Frequency- Inverse Document Frequency (TF-IDF) to vectorize such segments (submodule \scircle{1B} in Figure~\ref{fig:main-arch}). To answer why TF-IDF is effective here, we compare five different embedding techniques--Sentence BERT (SBERT), Doc2Vec, Universal Sentence Encoder (USE), Word2Vec, and TF-IDF~\cite{reimers-gurevych-2019-sentence,pmlr-v32-le14,cer-etal-2018-universal,mikolov2013efficient,sparck1972statistical}. Among them, TF-IDF appears to be most effective (see details in~\sectionref{subsec:embed}).

Considering each segment to be a document, formally for a term $t$ found in a document $d \in D$:
\[ tf(t,d) = \frac{f_{t,d}}{\sum_{t'\in d}f_{t',d}}\]
\vspace{-0.25cm}
\[ idf(t, D) = -\log P(t|D) = \log{\frac{n}{\sum{\mathbbm{1}_{(d\in D: t\in d)}}}}\]
\vspace{-0.15cm}
\[ tfidf(t,d, D) = tf(t,d) \cdot idf(t,D)\]

Here $f_{t,d}$ denotes the raw frequency of term $t$ in document $d$. In our problem, $d$ represents a text segment, and $D$ represents the corpus. 
TF-IDF maps the text segments $d$ in a $k$-dimensional latent space $\mathcal{X}$-
\vspace{-0.15cm}
\[ \phi: \mathcal{S} \rightarrow \mathcal{X} \quad d \in \mathcal{S}\]
In the next step, we use the cosine similarity score ($\psi$) to measure the similarity between each pair of TF-IDF vectors. 
For each vector $\myvec{x}_1$ and $\myvec{x}_2$:
\vspace{-0.15cm}
\begin{align*}
    \psi(\myvec{x}_1, \myvec{x}_2) &= \frac{\myvec{x}_1 \cdot \myvec{x}_2}{\left\lVert\myvec{x}_1\right\rVert \left\lVert\myvec{x}_2\right\rVert}
    = \frac{\sum_{i=1}^k\myvec{x}_{1i} \myvec{x}_{2i}}{\sqrt{\sum_{i=1}^k\myvec{x}_{1i}^2 \sum_{i=1}^k\myvec{x}_{2i}^2}}
\end{align*} 
We generate the symmetric matrix of all pair similarity scores. Now, from this symmetric matrix (Figure~\ref{fig:heatmaps}), we take the lower triangular half and remove the datapoints that have $\psi < \psi_{min}$ and $\psi > \psi_{max}$. 
We consider the following cases when choosing values for $\psi_{min}$ and $\psi_{max}$:

\begin{center}
\begin{figure}[ht]%
   \centering
   \subfloat[\centering]{{\includegraphics[width=0.53\linewidth]{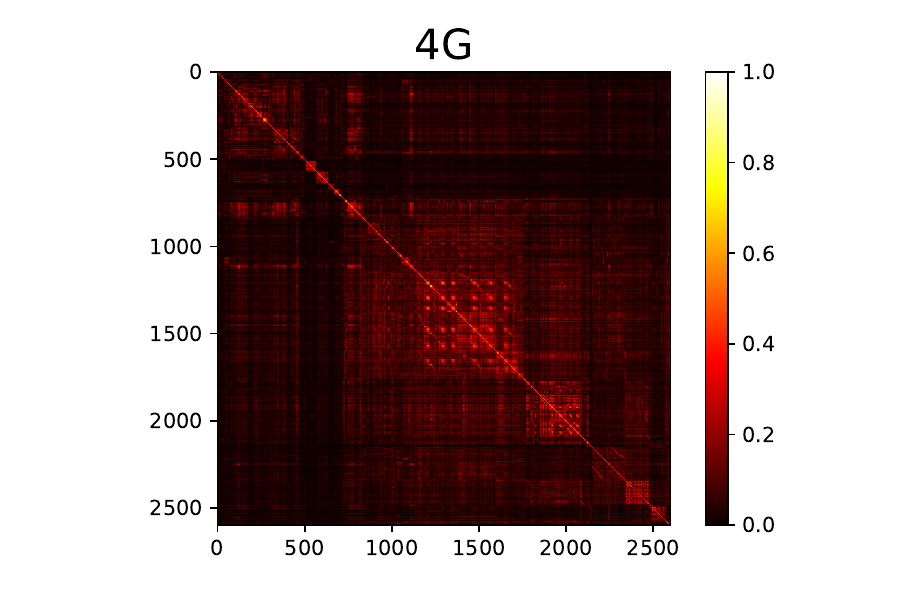} }}%
   \subfloat[\centering]{{\includegraphics[width=0.53\linewidth]{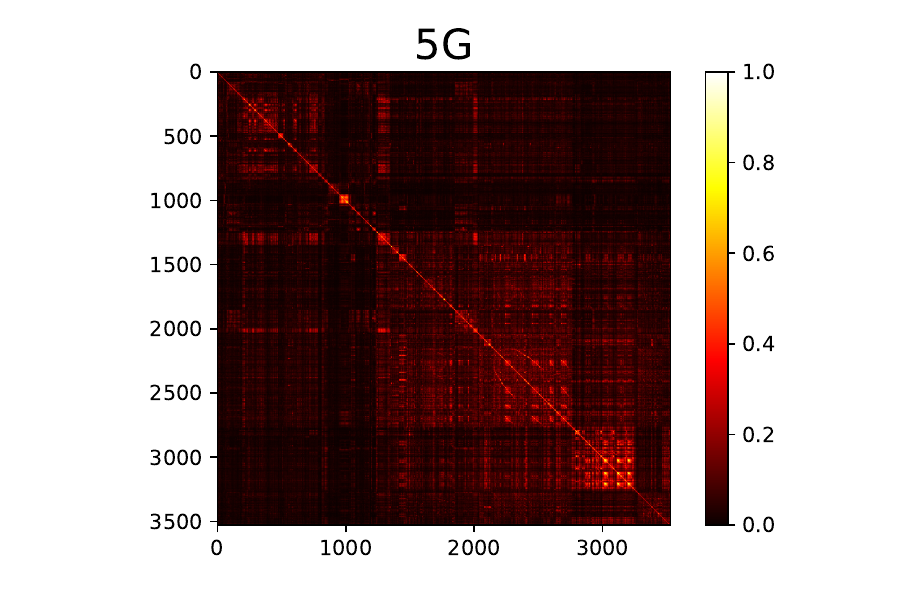} }}%
   \caption{Heatmap of the similarity matrices from 4G and 5G. The x and y axis represent the index of text segments extracted from the specifications. A brighter cell in the matrix represents more similarity.}%
   \label{fig:heatmaps}%
   \vspace{-0.8cm}
\end{figure}
\end{center}

\vspace{-0.3cm}
\begin{itemize}
\setlength\itemsep{-0.1em}
    \item When $\psi_{min}$ is too small, the segments are essentially describing two totally different events with a few words matched in the protocol specification. For a refined dataset, $\psi_{min}$ should be kept relatively high. 
    \item Using a $\psi_{max}$ helps to filter out segments where two text segments have very small changes in vocabularies, i.e., synonymous words or changes of articles ("a" instead of "the", pronouns instead of proper noun phrases, etc.). These segment pairs state the same event with (almost) exact description.
\end{itemize}
Further evaluation and discussion for the choice of $\psi$ are in section~\sectionref{sec:evaluation}.

This gives us a very useful set of filtered PoS. For supervised training, we go through another round of filtering to remove a few common PoS that are generated from the most common sentences in the documents. 

\subsection{Protocol Language Entailment Annotation}
We now discuss the hierarchical annotations of our dataset. The key challenge in finding inconsistencies in 4G/5G protocols is that there are no ground truth labels for model training. Thus, any supervised training task is hindered. We address this challenge by multi-phase training where the $0$th phase ensures the general capability of the model in distinguishing what an entailment or contradiction is, and the next phases are 
to precisely capture what sort of disparity is present in our actual cellular dataset.

The Stanford Natural Language Inference (SNLI) dataset contains datapoints with 3 unique annotations, namely, \textit{entailment, conflict}, and \textit{neutral}~\cite{snli:emnlp2015}. Since our model is first fine-tuned on this dataset (submodule \scircle{1C}) as a $0$th phase, we keep the same annotations to train the model(s) for subsequent phases. To characterize all scenarios, we consider \numlabel cases:
\begin{enumerate}[font={\bfseries},label={\arabic*:}]
\setlength\itemsep{-0.1em}
    \vspace{-0.05cm}\item $t_1 = t_2$: $t_1$ is consistent with $t_2$
    \item $t_1 \neq t_2$: $t_1$ is inconsistent with $t_2$
    \item $t_1 \otimes t_2$: $t_1$ is not related to $t_2$
    \item $t_1 \rightarrow t_2$: $t_1$ is related to $t_2$. $t_1$ happens before $t_2$
    \item $t_1 \leftarrow t_2$: $t_1$ is related to $t_2$. $t_2$ happens before $t_1$
    \item $t_1 \sqsupset t_2$: $t_1$ is related to $t_2$. $t_1$ contains more/detailed information than $t_2$
    \item $t_1 \sqsubset t_2$: $t_1$ is related to $t_2$. $t_2$ contains more/detailed information than $t_1$
\end{enumerate}

We argue that cases 1, 4, and 5 fall under entailment. Cases 2, 6, and 7 fall under contradiction. Case 3 maps onto the neutral label in the NLI task. For the contradiction class, cases 6 and 7 are difficult to perceive. We discuss this with a possible scenario. Suppose in a scenario both $t_1$ and $t_2$ independently describe the actions after an \packet{attach\_reject} is received by the UE with some specific EMM cause. Now, $t_1$ describes both the EPS status update and the security context clearing. In contrast,  $t_2$ describes only the  
process of updating the EPS status but does not mention anything about security context clearing. This will fall under case 6. This is considered as an inconsistency because the missing security context clearing in $t_2$ might create confusion in the implementation design and later on cause serious security issues.
\subsection{NLI Adaptation}

At each phase, our annotated protocol dataset contains \numtrainour PoS (in total 450 for three phases), whereas the SNLI dataset contains 570k English sentence pairs, which are $\sim$1266 times higher than the number of PoS in our dataset. Achieving even a reasonable performance is difficult if only this small dataset is used for transformer-based supervised learning, which contains millions of learnable parameters. Thus we first train transformer $\mathcal{T}$ on the SNLI dataset. This, in turn, produces a model $\mathcal{T}'$ that is ready to understand the difficult representation of conflicting statements. We then train $\mathcal{T}'$ on our \numtrainour data points to obtain $\mathcal{T}''_i$ for each 
fine-tune phase $i$, and finally test it on the large test dataset containing~\numtestour{}PoS for 4G and~\numtestournr{}for 5G.  Formally, the classification task can be defined as follows. For input $x$,
\begin{align*}
     h_{\mathcal{T}} &= \mathcal{T}(x) \\
 h_{fc} &= W_{fc}*h_{\mathcal{T}} + b_{fc} \\
 h_{drop} &= \text{\layer{Dropout}}(h_{fc}, p_d) \\
 \hat{y} &= \text{\layer{Softmax}}(h_{drop})
\end{align*}
where $\hat{y}$ denotes the predicted label. $W_{f_c}$ denotes the weights of 
the fully connected layer with $b_{f_c}$ being the bias. $p_d$ is the dropout rate. The overall model can be represented as:
\[\hat{y} = \text{\layer{Softmax}}(\text{\layer{Dropout}}(W_{fc}*\mathcal{T}(x) + b_{fc}, p_d))\]


\subsection{Few Shot Learning Optimization}
We emphasize that the straightforward end-to-end training is less interpretable and not robust at all. Since we have no ground truth validation set, in order to ensure the correctness of the learning phase, we must ensure that the learning process is robust and interpretable as much as possible. Therefore, we leverage two key techniques here over the end-to-end training:

\noindent \wcircle{1} \textbf{Transform \& Ensemble.} In this method, we use majority voting of multiple transformer models. 
We take BERT, RoBERTa, and XLNet for this approach~\cite{DBLP:journals/corr/abs-1810-04805,DBLP:journals/corr/abs-1907-11692,yang2019xlnet}. Note that we complete pre-training of these models before using them here. We call this ensemble model~\sysmodel{}. In \sectionref{sec:evaluation}, we discuss how this approach improves the confidence of the framework.

\noindent \wcircle{2} \textbf{Inconsistency-aware penalty.} As discussed earlier, we have a very small expert-annotated dataset, worse yet, with even fewer datapoints under the inconsistency class. 
In this method, we modify the standard cross entropy loss to penalize for wrong prediction on “inconsistent” labels. We use weighted cross-entropy loss:
\begin{align*}
   & \mathcal{L}_{wce} = \sum_{i=1}^{m}  {\hat{w}_y y_i\log{\hat{y}_i}}, \quad \text{where } w_y = \frac{N}{n_i} \text{, }\hat{w}_y = \frac{w_i}{\sum_{j}^{C}w_j} 
\vspace{-.2cm}
\end{align*}

Here, $N$ is the total number of training PoS, $n_i$ is the number of training PoS for class $i$, and $C$ is the total number of classes.
\noindent This assigns a higher penalty for misclassifying labels that have fewer PoS (in our case, the inconsistency class) than other classes.



%% file: sections/implementation.tex
\section{Implementation}\label{sec:implementation}
In this section, we discuss the implementation details of \system{}.

\noindent \textbf{Computational Hardware.}
For pre-training, we use about 2500 computing units from Google Colab pro+, which is equipped with Nvidia A100 GPUs. Multi-phase fine-tuning is performed on a computing server equipped with a 64-core CPU, 3x RTX 3090Tis, and 128GB RAM. Pre-training the base version of each model takes about 4 days in total. The uninformed fine-tuning on SNLI takes about 8-12 hours for each model, and for the informed multi-phase, each model requires different times, ranging between 10-15 minutes for each phase.

\noindent \textbf{Pre-training.}\label{subsubsec:pt}
We use BERT-base, RoBERTa-base, and XLNET-base as our candidate models. Each of these models is first pre-trained on the specification corpus. BERT-base and RoBERTa-base work on the Masked Language Model (MLM) objective and XLNET works on the Permutation Language Model (PLM) objective. The first two are pre-trained for $5$ epochs, and since XLNET-base is much larger in size, we pre-train it for $1$ epoch.

\noindent \textbf{Fine-tuning.}\label{subsubsec:ft}
 We use $k=3$ for the $k$-phase training. In the $0$th phase with the SNLI dataset, we keep the learning rate to be $3\times10^{-5}$ and batch size $32$. The models are trained for $8$ epochs. For all the consecutive fine-tuning phases, we observe that learning rate $2\times10^{-5}$ and batch size $32$ work better. The Adam optimizer with $\epsilon=10^{-8}$ has been used.
To subsidize the low training data, we add $\frac{N}{10}$ synthetic PoS at each phase. We augment the dataset with Easy Data Augmentation (EDA) to gain these synthetic PoS~\cite{wei-zou-2019-eda}. 
For both pre-training and fine-tuning, Huggingface transformers standard pipeline based on PyTorch has been used~\cite{hugg}. 

To validate the inconsistencies in UE implementation, we prepared Software Defined Radio (SDR)-based testbeds. For details of the testbed preparation, we refer the reader to~\cite{cellularlint}. Table~\ref{tab:ld} lists the devices we tested.


%% file: sections/evaluation.tex
\section{Evaluation}\label{sec:evaluation}
In this section, we evaluate the performance of \system{} from multiple perspectives--starting from ML training performance, identifying inconsistent descriptions of different levels, and ending with security implications. On a high level, we aim to answer the following research questions:

\begin{figure}[ht]%
   \centering
   \subfloat[\centering]{{\includegraphics[width=0.35\textwidth]{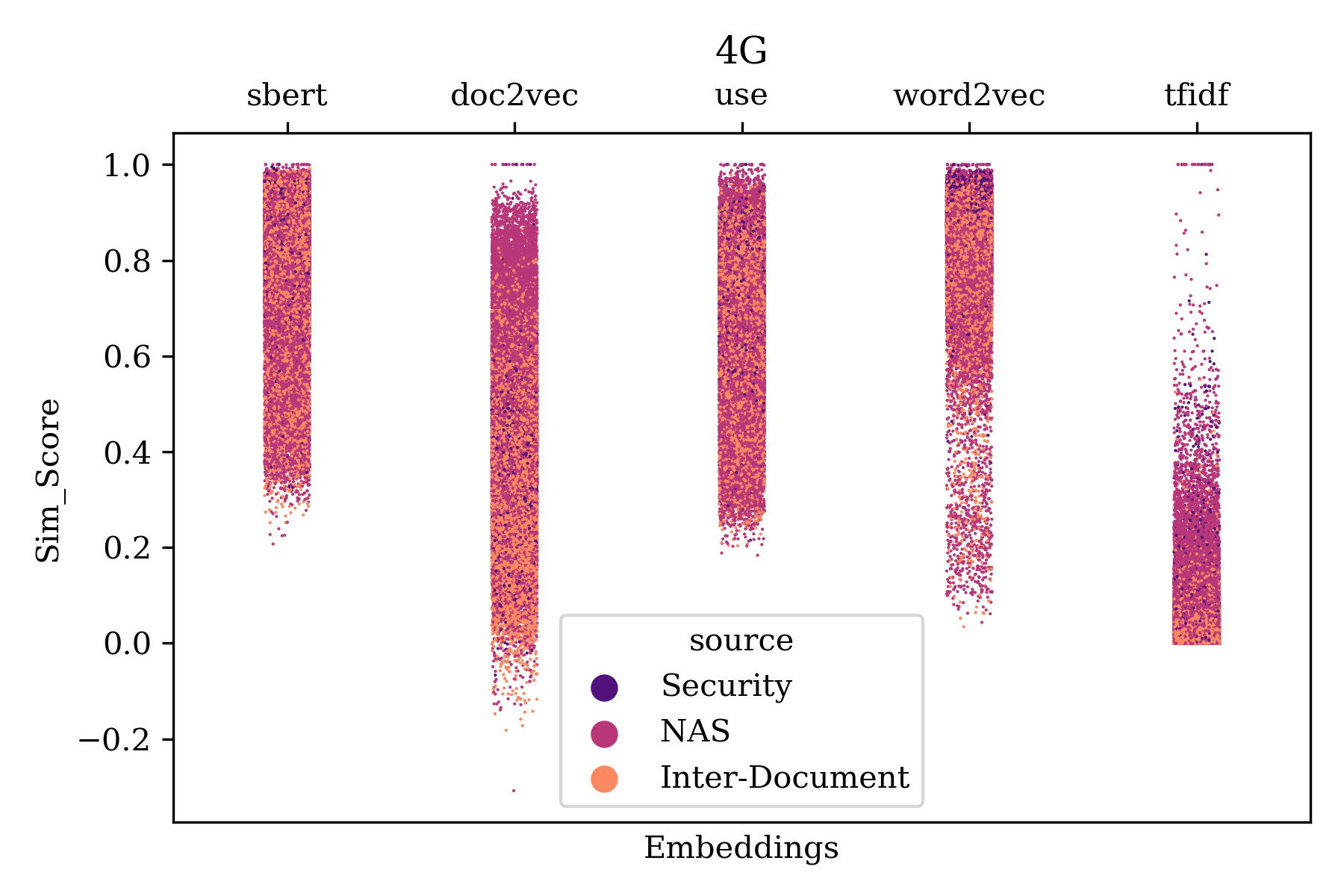} }}%
   \qquad
   \subfloat[\centering]{{\includegraphics[width=0.35\textwidth]{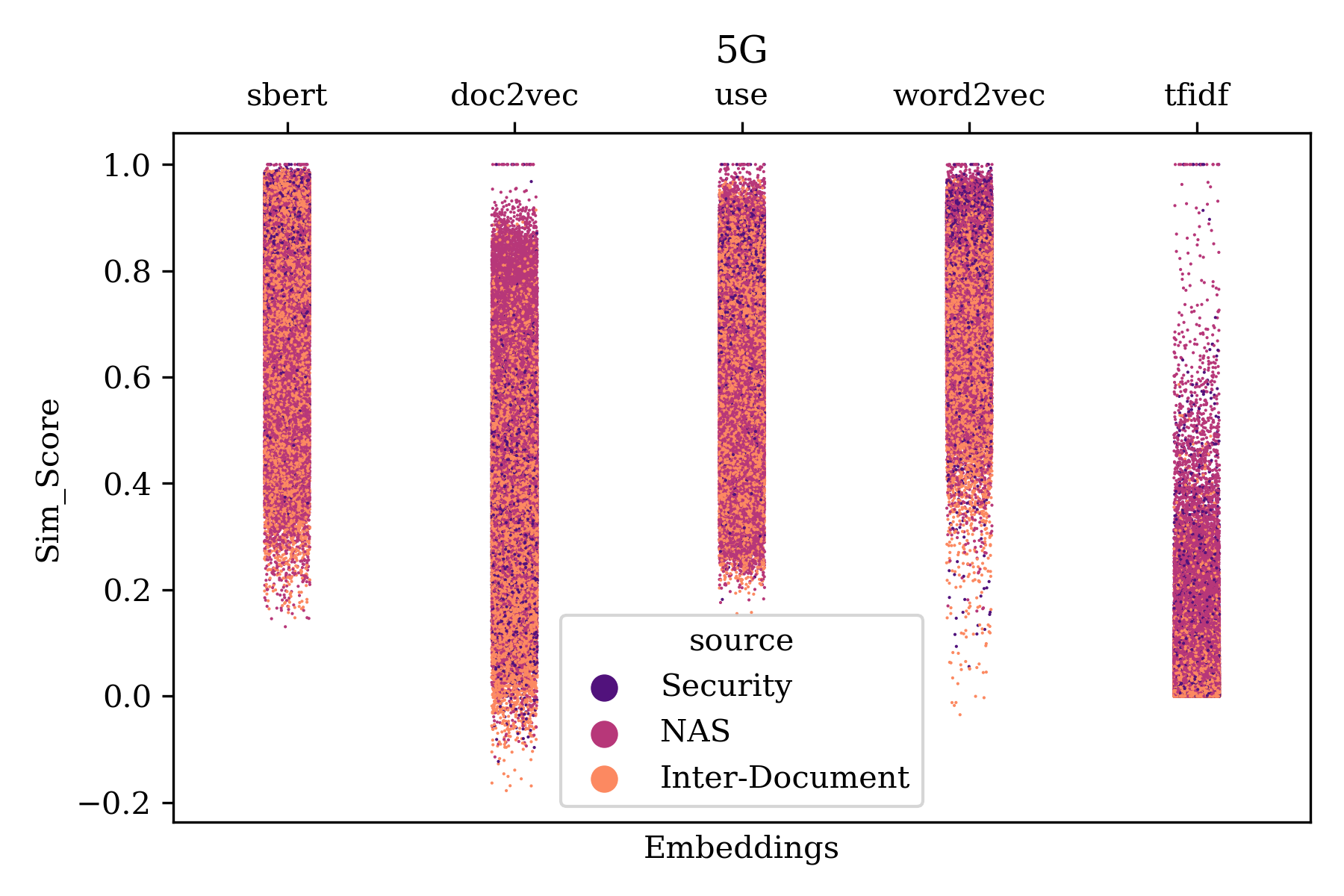} }}%
   \caption{Embedding comparisons. Only 0.8\% data were randomly sampled to generate the comparison for better visualization}%
   
   \label{fig:embedding}%
\end{figure}



\setlist{nosep} 
\begin{itemize}
\setlength\itemsep{-0.1em}
    \item \textit{RQ1. } How does the choice of embedding affect the search space contraction?
    \item \textit{RQ2. }What is the performance of~\sysmodel{} for inconsistency classification and how do the optimization techniques affect the performance?
    \item \textit{RQ3. }What are the issues found by~\system{} from specification documents and what are the security implications?
    
    \item \textit{RQ4. } how does this methodology compare to other related approaches in terms of the final issues found? 

    
\end{itemize}

\subsection{RQ1. Effect of embedding choice}\label{subsec:embed}
To benefit from the search space shrinkage through similarity measures, we identify the most effective embedding method for our dataset. 
Figure~\ref{fig:embedding} presents the comparison of embeddings based on the distribution of $\psi$. For an embedding method $\varepsilon$, each point in the figure represents the $\psi$ value of the PoS when $\varepsilon$ is used. For better visualization, only 0.8\% randomly sampled PoS are selected here. 
 Recall that our method filters out irrelevant pairs and only considers the highly similar (syntactically) segments for further stages. Among the 5 different embeddings, namely SBERT, Doc2Vec, USE, Word2Vec, and TF-IDF, TF-IDF gives the most sparse representation on the high similarity span. It is evident from the comparison that SBERT, Word2Vec, and USE cannot be utilized to reduce the search space 
 due to their very dense representation on the high similarity region. Doc2Vec shows some sparsity above $\psi=0.9$; however, the amount of PoS above that is very low. Following these results, we choose TF-IDF as our embedding technique for filtering.

\subsubsection{PoS Filtration}
For PoS filtering we set $\psi_{min} = 0.65$ for 4G and $\psi_{min} = 0.7$ for 5G, and $\psi_{max} = 0.99$ for both datasets. For 4G, we initially extracted 890 and 120 segments from NAS and security specifications, respectively. After sub-event 
driven quantization, we found 2599 unique segments for 4G. Considering all possible matches would give us $\sim6.75$M  datapoints, whereas our fine-grained approach limits the number of datapoints to only $1881$, 
which is $\sim 3591x$ lower. For 5G, $846$ and $329$ segments were extracted from NAS and security. After sub-event realization, we found $3529$ unique sequences for 5G. Considering all possible matches would give us $\sim12.45$M  datapoints whereas our fine-grained approach limits it to only $2541$ datapoints, which is $\sim4901x$ lower.

\begin{center}
    \begin{table*}[htbp]
  \centering
  \renewcommand{\arraystretch}{1}
	\fontsize{8}{8}\selectfont
 \caption{Precision, Recall, F1 Score, and Accuracy for the models at each phase.  \textit{Italic} represents the best performance in each phase. \textbf{Bold} represents the best overall. ${\text{Model}}_p$ denotes that the model has been \textit{pretrained}.}
  \begin{tabular}{cccccccccc}
   \hline
    \rule{0pt}{2ex}\multirow{2}{*}{Model} & \multicolumn{4}{c}{Phase 0} & & \multicolumn{4}{c}{Phase 1}\\
    \cline{2-5}\cline{7-10}
   \rule{0pt}{2ex} & Precision & Recall & F1 Score & Accuracy & & Precision & Recall & F1 Score & Accuracy\\
    \hline
    \rule{0pt}{2ex}$\text{BERT}_p$ & 0.4606 & 0.4536 &	0.4547& 0.5133 & & 0.6125	& 0.6222 & 0.6114 & 0.68 \\
    \hline
    \rule{0pt}{2ex}$\text{RoBERTa}_p$ & \textit{0.5033}	& 0.5027& \textit{0.5017} & \textit{0.5667} & & \textit{0.6558} & \textit{0.6606} & \textit{0.6515} & \textit{0.7133} \\
    \hline
   \rule{0pt}{2ex} $\text{XLNet}_p$ & 0.4718 & 0.4748 & 0.4695 & 0.5467 & & 0.5914 & 0.6041 & 0.591 & 0.6667  \\
    \hline
   \rule{0pt}{2ex} $\text{\sysmodel{}}_p$ & 0.5002 & \textit{0.5029} & 0.498 & \textit{0.5667} & & 0.6167 & 0.6343 & 0.6171 & 0.6867 \\
    \hline\hline
    \rule{0pt}{2ex}\multirow{2}{*}{Model} & \multicolumn{4}{c}{Phase 2} & &  \multicolumn{4}{c}{Phase 3}\\
    \cline{2-5}\cline{7-10}
    \rule{0pt}{2ex}& Precision & Recall & F1 Score & Accuracy & &  Precision & Recall & F1 Score & Accuracy\\
    \hline
    \rule{0pt}{2ex}$\text{BERT}_p$ & 0.6859 & 0.7026 & 0.689 & 0.7467 & &  0.7473 & 0.7787 & 0.7562 & 0.7933\\
    \hline
    \rule{0pt}{2ex}$\text{RoBERTa}_p$ & 0.6967 & 0.7062 & 0.6947 & 0.7533 & & 0.7788 & 0.7695 & 0.7676 & 0.8133 \\
    \hline
    \rule{0pt}{2ex}$\text{XLNet}_p$ & 0.6534 & 0.6669 & 0.6552 & 0.72 & &  0.6986 & 0.7097 & 0.6901 & 0.7533 \\
    \hline
    \rule{0pt}{2ex}$\text{\sysmodel{}}_p$ & \textit{0.7262} & \textit{0.7404} & \textit{0.7249} & \textit{0.78} & & \textit{\textbf{0.7871}} & \textit{\textbf{0.8148}} & \textit{\textbf{0.7962}} & \textit{\textbf{0.8267}} \\
    \hline
  \end{tabular}
  \label{tab:scores}
\end{table*}
\end{center}
\begin{figure*}[ht]
      \centering
      \begin{subfigure}[b]{0.24\textwidth}
          \centering
          \includegraphics[width=\textwidth]{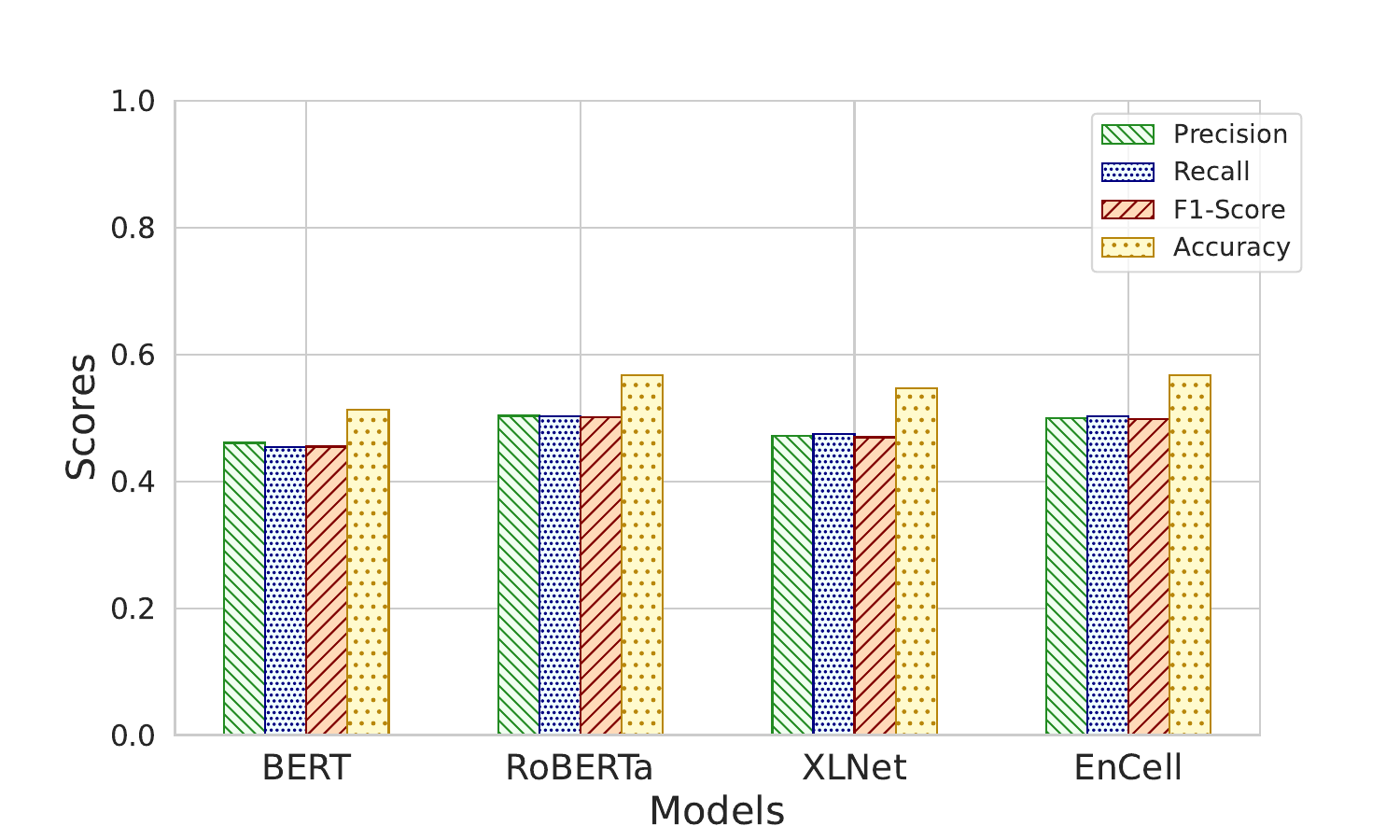}
          \caption{Phase 0}
          \label{fig:bar1}
      \end{subfigure}
      \begin{subfigure}[b]{0.24\textwidth}
          \centering
          \includegraphics[width=\textwidth]{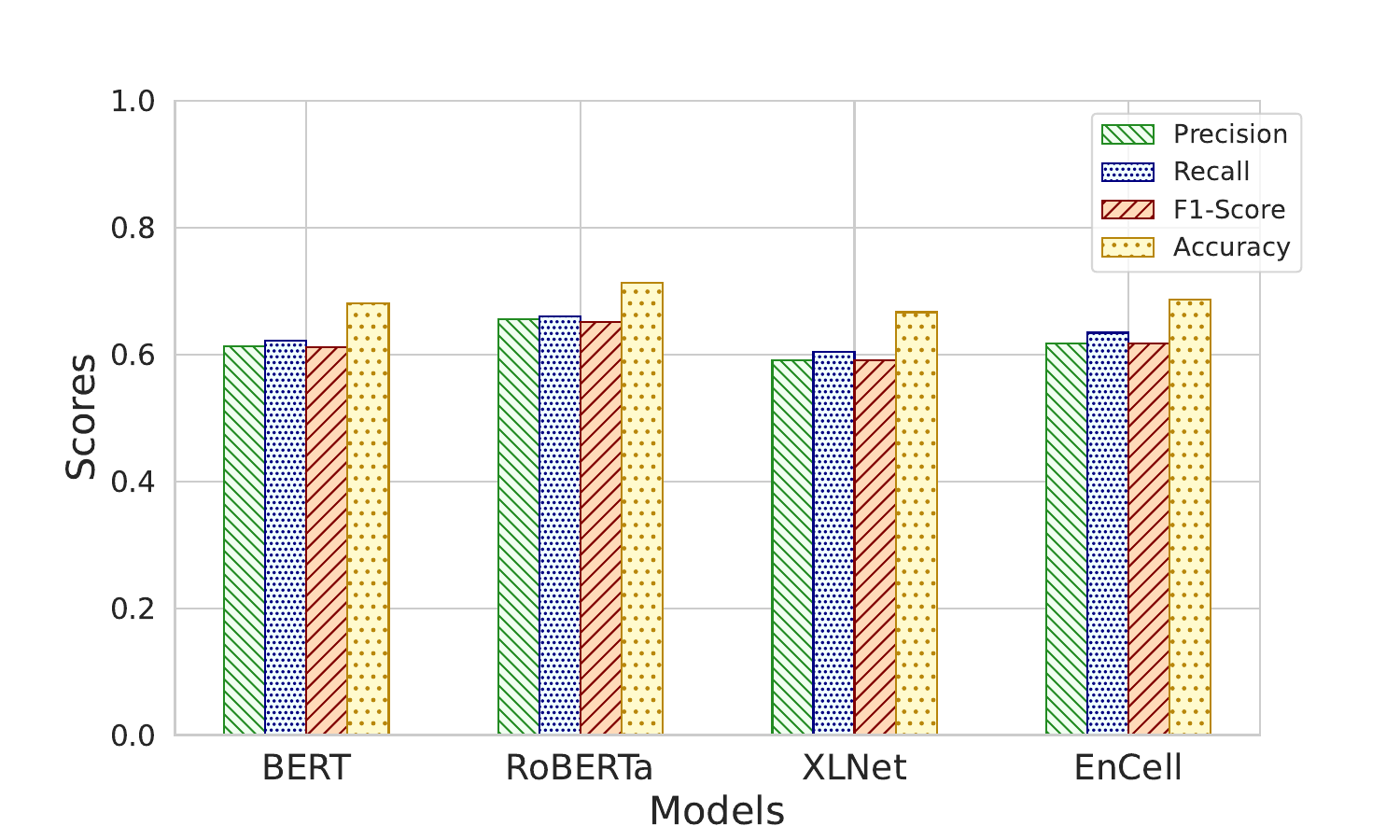}
          \caption{Phase 1}
          \label{fig:bar2}
      \end{subfigure}
      \begin{subfigure}[b]{0.24\textwidth}
          \centering
          \includegraphics[width=\textwidth]{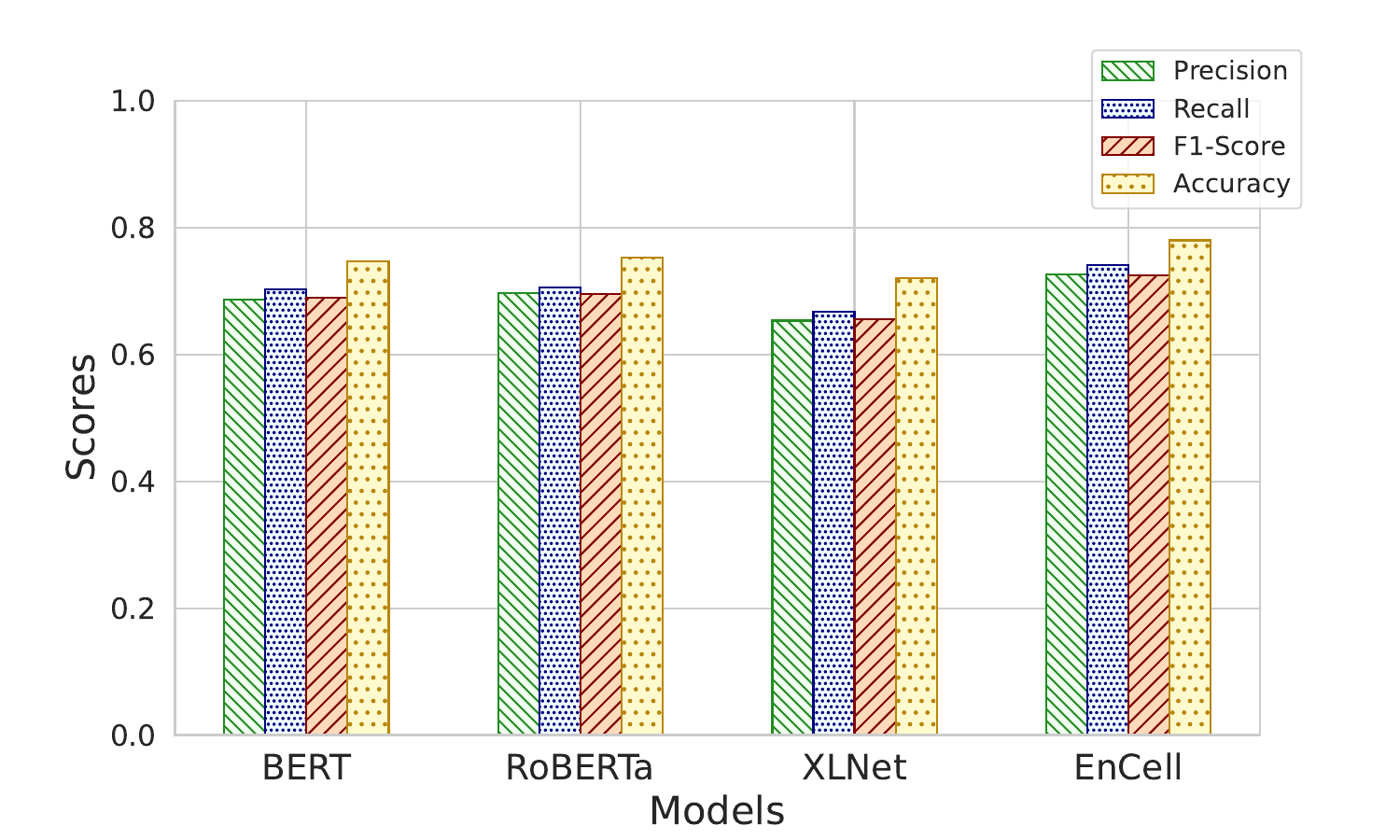}
          \caption{Phase 2}
          \label{fig:bar3}
      \end{subfigure}
      \begin{subfigure}[b]{0.24\textwidth}
          \centering
          \includegraphics[width=\textwidth]{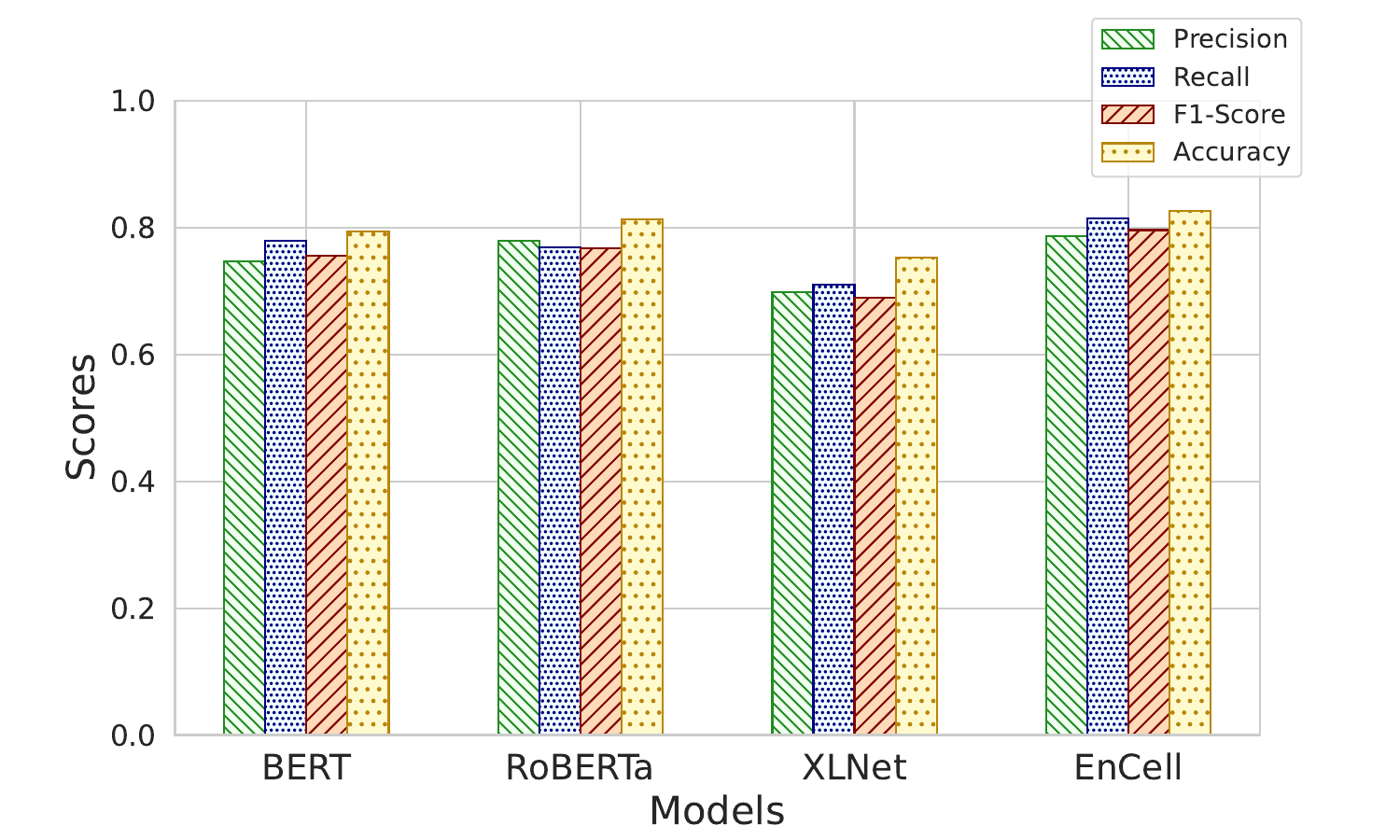}
          \caption{Phase 3}
          \label{fig:bar4}
      \end{subfigure}
         \caption{Performance metrics for different models used by \system{}}
         \label{fig:all score}
 \end{figure*}
\subsection{RQ2. \sysmodel{} performance}
To evaluate how effectively the models and their ensemble \sysmodel{} adapt to the domain and classify inconsistencies, we evaluate them at each phase of the fine-tuning. 
Here, we show the precision, recall, F1-score, and accuracy of the models at each phase (Table~\ref{tab:scores} and Figure~\ref{fig:all score}). Note that these are the $k=0, \dots, 3$ phases of training we mentioned in submodule \scircle{1E} of our method.
Before the evaluation, all these models are pre-trained on the cellular network specification corpus and fine-tuned on the general-purpose SNLI dataset. Moreover, here all the models use weighted cross-entropy loss in all phases.

We observe that from phase 0 to phase 3, the F1 score improves about $29\%$ for \sysmodel{}. RoBERTa performs better in the first two phases while \sysmodel{} does better in phase 3. This is expected as RoBERTa performs the best among the available pre-trained transformers on general NLI tasks. Once BERT and XLNet go through two phases of training, they start to perform in the vicinity of RoBERTa. This coherence of performance motivates the use of the ensemble approach in our task. Interestingly, although the accuracy of phase-3 RoBERTa ($81.33\%$) and \sysmodel{} ($82.67\%$) is similar, we observe approximately $2.9\%$ improvement of \sysmodel{} over the F1 score of RoBERTa and $3.4\%$ improvement over the best recall score provided by BERT. This emphasizes the need to use \sysmodel{}  in an uncertain problem setting such as ours.  

\noindent\textbf{Effect of inconsistency-aware penalty.}
To understand the effectiveness of the data distribution-based weighted cross-entropy loss,
we compare what happens when the general cross-entropy loss is used at phase 1 training and when the inconsistency-aware penalty is used. 
When general cross-entropy loss is used, we see only $3-5\%$ improvement on F1 score of our model. In contrast, the inconsistency-aware loss improves the F1 score of about $6-13\%$ at phase 1. This clearly indicates the usefulness of the penalty in our setting.

\subsection{RQ3. Issues found by \system{}}
\system{} found a total of \numissuesemantic{} inconsistent PoS from 4G and 5G specifications. We observe that among them, there are certain scenarios that, if taken in a broader context, do not contradict. Yet the model predicts them as inconsistencies since it cannot always understand the large event at a single inference. Note that we do not consider them as false positives; provided only the context in the PoS, we can see that they are semantically conflicting. However, the additional context provided in the original specification shows that the larger context is different. In some cases, the larger context is defined 3-4 pages earlier than the original PoS.  


This is a fundamental challenge for NLP-based conflict detection for cellular network protocol specifications. 
Therefore, we manually filter out such 
apparent inconsistencies and group the PoS into different categories. After filtering, we found a total of \numissuesclean{} PoS that may affect design choices. Table~\ref{tab:cat breakdown} shows the categorical breakdown of inconsistent pairs found by \system{}. To do this manual filtration, we spent 6 human hours of time and found around 16\% of issues detected by \system{} had to be filtered out due to broader context.

\begin{center}
\begin{table}[H]
  \centering
  \caption{Conflicting statements breakdown by category}
    \captionsetup{font={small}}
 	\renewcommand{\arraystretch}{1}
 	\fontsize{7}{7}\selectfont
  \begin{tabular}{c|cc}
    \hline
    & Category & Count \\
    \cline{2-3}
    \rule{0pt}{2ex}\multirow{7}{*}{4G} & Integrity \& ciphering & 26\\
    & Security context handling & 2\\
    & Bearer context & 1 \\
    & GUTI related & 41 \\
    & State transition & 12 \\
    & Counters & 6\\
    & Misc & 3 \\
    \hline
    \rule{0pt}{2ex}\multirow{11}{*}{5G} & Security key generation \& handling & 3\\
    & Integrity \& ciphering & 3 \\
    & Timers & 7 \\
    & State transition & 2 \\
    & Paging & 3 \\
    & PLMN handling & 3 \\
    & GUTI related & 3 \\
    & QoS rules & 5 \\
    & PDU session establishment & 30 \\
    & Counters & 2 \\
    & Misc & 5 \\
    \hline
    \rule{0pt}{2ex} & Total & \numissuesclean{}
  \end{tabular}
  \label{tab:cat breakdown}
  \vspace{-0.9cm}
\end{table}
\end{center}

\subsubsection{Investigation of the issues}
\label{subsec:impact}
To further investigate how the discovered issues may impact design choices, we examine them against open-source 4G and 5G implementations from srsRAN (v23.04.1), Open5GS (v 2.6.6), and OpenAirInterface (v2.0.0)~\cite{srsRAN,open5GS,openAirInterface}.
We look into both the UE and Core-Network implementations of these open-source implementations, which are widely used to analyze cellular network security~\cite{9519388,279972,10.1145/3460120.3485388}. Furthermore, in case the inconsistencies can lead to security, privacy, availability or interoperability issues, we manually investigate and design exploits for the scenarios.

\begin{center}

\def\arraystretch{1}
\begin{table}[ht]
	\centering
    \caption{Conflicting behaviors discovered from 4G Implementations 
    }
    \captionsetup{font={small}}
    \caption*{\cleft = 1st type, \cright = 2nd type, \cno = Did not implement, \unsafe{Red} :  Unsafe choice, * : partially implemented/related implementation found}
	\renewcommand{\arraystretch}{1}
	\fontsize{7}{7}\selectfont
\begin{tabular}{p{0.12\linewidth}p{0.55\linewidth}p{0.08\linewidth}p{0.08\linewidth}}
\rule{0pt}{2ex}Finding & Description & srsRAN & OAI \\ [0.5ex] 
 \hline\hline

 \hline
 \rule{0pt}{2ex} 1 & NAS message ciphering & \unsafe{\cleft} & \unsafe{\cleft} \\
 \hline
 \rule{0pt}{2ex} 2 & Condition over Integrity Check & \unsafe{\cright} & \unsafe{\cright} \\
 \hline
 \rule{0pt}{2ex} 3 & Integrity check failure & \unsafe{\cright} & \cleft \\
 \hline
 \rule{0pt}{2ex} 5 & GUTI Deletion & \cleft* & \cno \\
 \hline
 
 \stepcounter{lteic}
 \rule{0pt}{2ex}7.1 & EMM deregistered sub-state transition & \unsafe{\cright}* & \cno \\
 \hline
 \stepcounter{lteic}
 \rule{0pt}{2ex}7.2 & EMM registered sub-state transition & \cleft & \cno \\
 \hline
 \rule{0pt}{2ex} 9~\cite{cellularlint} & Timer expiry as precondition & \cleft* & \cleft \\
 \hline
 \rule{0pt}{2ex}10~\cite{cellularlint} & TAU/Service attempt counter usage & \unsafe{\cright} & \cno \\
 \hline
 
\end{tabular}
\vspace{-0.5cm}
\label{tab:4g}
\vspace{-0.5cm}
\end{table}
\vspace{-0.8cm}
\end{center}

\begin{center}

\def\arraystretch{1}
\begin{table}[ht]
	\centering
    \caption{Conflicting behaviors discovered from 5G Implementations
    }
    \captionsetup{font={small}}
    \caption*{\cleft : 1st type, \cright : 2nd type, \cno : Did not implement, \unsafe{Red} :  Unsafe choice, A: Additional}
	\renewcommand{\arraystretch}{1}
	\fontsize{7}{7}\selectfont
\begin{tabular}{p{0.1\linewidth}p{0.38\linewidth}p{0.1\linewidth}p{0.08\linewidth}p{0.08\linewidth}}
\rule{0pt}{2ex}Finding & Description & Open5GS (Core) & SrsRAN (UE) & OAI  \\ [0.5ex] 
 \hline\hline
 \rule{0pt}{2ex} 1 & NAS message ciphering & \unsafe{\cleft} & \unsafe{\cleft} & \unsafe{\cleft}\\
 \hline
 \rule{0pt}{2ex} 2 & Condition over integrity check & \unsafe{\cright} & \unsafe{\cright} & \unsafe{\cright} \\
 \hline
 \rule{0pt}{2ex} 4 & NCC reusage & \unsafe{\cright} & \unsafe{\cright} & \unsafe{\cright} \\
 \hline
 \rule{0pt}{2ex}5.1 & 4G-GUTI deletion & \cno & \cleft* & \cno \\
 \hline
 \stepcounter{nric}
 \rule{0pt}{2ex}5.2 & 5G-GUTI deletion & \cno & \cleft* & \cno \\
 \hline
 \rule{0pt}{2ex} 8~\cite{cellularlint} & Reset attempt counter & \cno & \unsafe{\cright} & \cno \\
 \hline
 \rule{0pt}{2ex} A~\cite{cellularlint} & SUPI in auth\_response & \unsafe{\cright} & \cno & \cno \\
 \hline
\end{tabular}
\label{tab:5g}
\end{table}

\end{center}

Table~\ref{tab:4g} and \ref{tab:5g} report the result of our observation. Note that, even in open-source implementations a lot of functionalities are not implemented and scenarios related to protocol handovers could not be tested. To further validate the impact of the identified inconsistent descriptions, we perform end-to-end attacks in \numue COTS UE devices (Table~\ref{tab:ld}). 
Note that here as well we could not confirm issues related to network states as it is not possible to confirm the specific substate of the commercial devices. Furthermore, we could not execute any attacks on the commercial carrier side as it is prohibited by law. Nonetheless, we have reported all our findings to 3GPP and the vendors. We want to emphasize the ultimate aim of \system{} is to find inconsistent statements in the specifications, and the implementation testing (both open-source and commercial UEs) is done to further validate the impact conflicting statements can have on implementation design. Though we are inherently bound by the issues we can test, we take a \emph{best-effort} approach in validating the impacts. In the following sections, we go into our threat model and the details of the findings.

\noindent \textbf{Threat model.}
We consider the communication channel between the core network, base station, and UE to be under the influence of a Dolev-Yao~\cite{1056650} 
attacker. In this threat model, an adversary can generate/parse messages from scratch or modify any intercepted message while not having any cryptographic capabilities. Concretely, we consider three scenarios commonly assumed in 4G and 5G: Man-in-the-Middle (MitM), Fake Base Station (FBS), and signal injection attacks. The threat model is consistent with the existing approaches that consider real-world threats~\cite{236354,277258,10.1145/3460120.3485388}.   




\subsubsection{Security Enforcement}
\stepcounter{fdcounter}
\noindent\findingsty{\thefdcounter. }\textit{NAS message ciphering.}
After the completion of security mode procedures, the specifications suggest that, except for a few messages, all messages should be integrity-protected and ciphered. However, on two occasions (shown in Figure~\ref{lt:ciphering}, it is emphasized that all messages after the security procedure should be integrity-protected and ciphered. 
Failure to enforce integrity and ciphering is a serious design flaw and can leave implementations subject to vulnerabilities and interoperability issues. 

\begin{figure}[htbp]
\vspace{-0.5cm}
    \centering
   \noindent\begin{mdframed}[font=\scriptsize]
    $T_1$: From this time onward the UE shall cipher and integrity protect all NAS signalling messages with the selected NAS ciphering and NAS integrity algorithms.
    \\
    \noindent $T_2$: From this time onward, all NAS messages exchanged between the UE and the MME are sent integrity protected and \textbf{except for the messages} specified in clause 4.4.5, all NAS messages exchanged between the UE and the MME are sent ciphered
    \end{mdframed}
    \vspace{-0.5cm}
    \caption{Ciphering exceptions. $T_1$ is from section 5.4.3.3 of TS 24.301 and $T_2$ is from section 4.4.2.5 of TS 24.301}
    \label{lt:ciphering}
    
\end{figure}

\looseness = -1
\noindent \textbf{Inconsistency to exploit.} In the case of $T_1$, the specification states after the security context has been established, the UE shall cipher and integrity protect all NAS signaling messages with selected NAS ciphering and integrity algorithms. However, $T_1$ does not mention the scenarios where the UE accepts plain-text messages. On the contrary, $T_2$ is more lucid and explicitly mentions such scenarios. Due to this inconsistency of not explicitly mentioning exception cases, an implementation might accept unexpected plain-text messages even after the security context is established. Therefore, we design an attack based on this scenario.

\noindent \textbf{Attack.} 
For the attack, the adversary connects with a victim UE through a fake base station to send plain-text \packet{identity\_request} or \packet{authentication\_request}~\cite{277258}. Similarly, the attacker can overshadow downlink packets to create plain-text messages. The affected UE accepts, processes and responds to the messages. 

\looseness = -1
\noindent \textbf{Investigation.}
In open-source implementations, we found all implementations accepting plain-text \packet{authentication} and \packet{identity} requests even after the security context has been established. On the other hand, on commercial UEs, we found 4 UEs accepting and responding to such plain-text messages (Table~\ref{tab:uetest}). Interestingly, for a very recent UE (Google Pixel 7a, 2023),  plain-text \packet{authentication\_request} is accepted, whereas plain-text \packet{identity\_request} is not accepted. This is another interesting scenario where vendors try to implement different behaviors for different exception case messages.

\noindent \textbf{Impact.} 
The impact of accepting plain-text or integrity-failed NAS messages can be catastrophic and can be exploited by attackers to fingerprint users, traceability, and denial of service attacks. Note that DoLTEst~\cite{277258} also reports several UEs accepting unprotected messages after the security context has been established, albeit from the implementation testing perspective. We, on the other hand, show that this issue in different implementations can be traced to the inconsistent behavior defined in the standards.


\stepcounter{fdcounter}
\noindent\findingsty{\thefdcounter. }\textit{Condition over integrity check.}
On many occasions, the UE sets the counter for "SIM/USIM considered for GPRS/non-GPRS/5GS services" to implement a specific maximum value (Figure~\ref{lt:integ_check}). While this flexibility is acceptable and standard practice, the precondition to check the integrity of the received message is often neglected. 

\begin{figure}[htbp]
    \centering
    \vspace{-0.5cm}
\noindent\begin{mdframed}[font=\scriptsize]
$T_1$: The UE shall consider the USIM as invalid for EPS services and non-EPS services until switching off or the UICC containing the USIM is removed or the timer T3245 expires as described in clause 5.3.7a. Additionally, the UE shall delete the list of equivalent PLMNs and enter state EMM-DEREGISTERED.NO-IMSI. \textbf{If the message has been successfully integrity checked} by the NAS and the UE maintains a counter for "SIM/USIM considered invalid for GPRS services", then the UE shall set this counter to UE implementation-specific maximum value.
\\
\noindent $T_2$: The UE shall consider the USIM as invalid for EPS services until switching off or the UICC containing the USIM is removed or the timer T3245 expires as described in clause 5.3.7a. The UE shall delete the list of equivalent PLMNs and shall enter the state EMM-DEREGISTERED.NO-IMSI. If the UE maintains a counter for "SIM/USIM considered invalid for GPRS services", then the UE shall set this counter to UE implementation-specific maximum value.
\end{mdframed}
\vspace{-0.5cm}
    \caption{Condition over integrity check. $T_1$ is from section 5.5.1.2.5 and $T_2$ is from section 5.5.2.3.2 of TS 24.301.
    }
    \label{lt:integ_check}
\end{figure}

\noindent \textbf{Inconsistency to exploit.} In the case of $T_1$, the protocol specifies successful integrity checking by the NAS. However, in $T_2$, this integrity checking is skipped. These PoS are related to \packet{attach\_reject} and \packet{detach\_request}. However, we have also found instances of network-initiated \packet{detach\_request}, where there are inconsistencies. 
As the inconsistency is related to the integrity checking of messages, the attack steps and impact are the same as finding 1.

\noindent \textbf{Investigation.} In the case of open-source implementation, we see that all the open-source implementations in both 4G and 5G are vulnerable. However, in the case of commercial UEs, there were no instances of vulnerable behavior.

\stepcounter{fdcounter}
\noindent\findingsty{\thefdcounter.} \textit{Integrity check failure.}
We found two conflicting PoS where different statements are found for failed integrity checks of the control plane messages (Figure~\ref{lt:4.1}).  Following this PoS to 4G security specification, ultimately, we found three (one segment is common) instances of different statements.  Two of them clearly suggest that messages that have faulty MAC should be discarded, whereas the third one directs to a slightly flexible strategy suggesting the processing of certain messages even if they fail integrity checks. 

\begin{figure}[htbp]
    \centering
   \noindent\begin{mdframed}[font=\scriptsize]
    $T_1$: In case of failed integrity check (i.e. faulty or missing
MAC-I) is detected after the start of integrity protection,
the concerned message shall be discarded. This can happen on the UE side or on the eNB side
    \\
    \noindent $T_2$: In case of failed integrity check (i.e. faulty or missing NAS-MAC) is detected after the start of NAS integrity protection the concerned message shall be discarded \textbf{except for some NAS messages} specified in
    TS 24.301 [9]. For those exceptions the MME shall take the actions $\cdots$ NAS message with faulty or missing NAS-MAC
    \end{mdframed}
    \caption{Allowing exceptions for integrity failure. $T_1$ is from section 7.3.2 and $T_2$ is from section 8.1.1 of TS 33.401.}
    \label{lt:4.1}
\end{figure}

\noindent \textbf{Inconsistency to exploit.}
In $T_1$, whenever there are some exceptional NAS messages that can trigger further MME actions, even with failed integrity, the specification mentions that. However, in $T_2$ for RRC, it does not specify such exceptional cases, though such exceptional cases exist, for instance, when \packet{RRC\_connection\_resume} fails an integrity check, rather than just discarding, there are further steps. Through further manual analysis, we find these cases are not mentioned in the TS 33.501 (Security architecture and procedures for 5G Systems) specification but in TS 38.331 (RRC specification), which were not included in the scope of \system{}. In a practical scenario, an implementor would use the RRC specification first, which clearly states exceptional messages with failed integrity. But later on, when the security specifications would be taken into account, the implementor would find it confusing with a more strict description. There is a possibility that such inconsistency may not directly result in vulnerabilities. Still, it may cause differing implementations as both specifications cannot be logically taken into consideration at the same time. Hence, some implementations might not properly follow the integrity failing scenarios. 

\noindent \textbf{Investigation.} 
In open-source implementations, we found 1 implementation accepting integrity-failed messages. On the contrary, in commercial UEs, none of the UEs accept control-plane messages with failed integrity. However, in our investigation, we found another interesting behavior: 16 UEs dropped the connection after receiving an RRC packet with failed integrity, 
whereas 1 UE did not (Table~\ref{tab:uetest}). These inconsistent UE behaviors can be traced back to $T_2$, which does not properly specify the exception cases and what to do in such scenarios.

\stepcounter{fdcounter}
\noindent\findingsty{\thefdcounter.} \textit{NCC reusage. }
We found a conflicting PoS in the 5G Security specification where two different conditions are 
expressed for the validity of the Next hop Chaining Counter (NCC) (shown in Figure~\ref{lt:3.1}). NCC is used for cryptographic derivation of the AS security algorithms and, hence, is a very important identifier. One segment dictates that the NCC value has to be \emph{fresh} and previously \emph{unused} to be accepted. On the contrary, the other segment claims that the only condition for acceptance is that the NCC value has to be \emph{different}. 
\begin{figure}[htbp]
\vspace{-0.5cm}
    \centering
   \noindent\begin{mdframed}[font=\scriptsize]
    $T_1$: If the sent NCC value is \textbf{fresh} and belongs to an \textbf{unused pair} of {NCC, NH}, the gNB shall save the pair of \{NCC, NH\} in the current UE AS security context and shall delete the current AS key $K_{gNB}$.
    \\
    \noindent $T_2$: The UE shall take the received NCC value and save it as stored NCC $\cdots$ . If the stored NCC value is \textbf{different} from the NCC value associated with the current KgNB, the UE shall delete the current AS key $K_{gNB}$
    \end{mdframed}
    \vspace{-0.5cm}
    \caption{Underspecified action for NCC. Both $T_1$ and $T_2$ are from section 6.8.2.1.2 of TS 33.501}
    \vspace{-0.5cm}
    \label{lt:3.1}
    \vspace{-0.5cm}
\end{figure}


\begin{center}

\def\arraystretch{1}
\begin{table}[ht]
	\centering
    \caption{List of devices tested}
    \captionsetup{font={small}}
	\renewcommand{\arraystretch}{1}
	\fontsize{7}{7}\selectfont
\begin{tabular}{P{0.27\linewidth}P{0.08\linewidth}P{0.08\linewidth}P{0.4\linewidth}}
\rule{0pt}{2ex} Device & Version & Release Year & Baseband \\ [0.5ex] 
 \hline\hline

 \rule{0pt}{2ex} Google Pixel 7a & 13 & 2023 & 	Google Tensor G2  \\
 \hline

 \rule{0pt}{2ex} Samsung Galaxy S20 FE &  \multirow{ 2}{*}{10} &  \multirow{ 2}{*}{2020} & Qualcomm SM8250 Snapdragon 865 5G  \\
 \hline

 \rule{0pt}{2ex} HTC One E9 &  7.0 &  2015 & Mediatek MT6795M Helio X10  \\
 \hline
 \rule{0pt}{2ex} Huawei Y5 & 9.0 & 2019 & Mediatek MT6761 Helio A22  \\
 \hline
 
  \rule{0pt}{2ex} \multirow{ 2}{*}{Xiaomi Mi 11 Lite 5G} &  \multirow{ 2}{*}{11} &  \multirow{ 2}{*}{2021} & Qualcomm SM7350-AB Snapdragon 780G  \\
 \hline

   \rule{0pt}{2ex} \multirow{ 2}{*}{Motorola Edge 30 Pro} &  \multirow{ 2}{*}{12} &  \multirow{ 2}{*}{2022} & Qualcomm SM8450 Snapdragon 8 Gen  \\
 \hline

   \rule{0pt}{2ex} \multirow{ 2}{*}{OnePlus 9 Pro} &  \multirow{ 2}{*}{11} &  \multirow{ 2}{*}{2021} & Qualcomm SM8350 Snapdragon 888 5G   \\
 \hline

    \rule{0pt}{2ex} Honor 8X & 8.1 & 2018 & Kirin 710   \\
 \hline
    \rule{0pt}{2ex}\multirow{ 2}{*}{Apple iPhone 12 Pro}  & iOS 17.3 &  \multirow{ 2}{*}{2020} &  \multirow{ 2}{*}{Apple A14 Bionic}   \\
 \hline

\rule{0pt}{2ex}  \multirow{ 2}{*}{Google Pixel 3a} &  \multirow{ 2}{*}{9} &  \multirow{ 2}{*}{2019} & Qualcomm SDM670 Snapdragon 670   \\
 \hline
 
  \rule{0pt}{2ex} Samsung Galaxy A04  &  12 &  2022 & Mediatek MT6765 Helio P35   \\
 \hline



  \rule{0pt}{2ex}  \multirow{ 2}{*}{LG Velvet 5G}  &  \multirow{ 2}{*}{10} &  \multirow{ 2}{*}{2020} & Qualcomm SM7250 Snapdragon 765G 5G   \\
 \hline

  \rule{0pt}{2ex} \multirow{ 2}{*}{OnePlus 8T}  &  \multirow{ 2}{*}{11} &  \multirow{ 2}{*}{2020} & Qualcomm SM8250 Snapdragon 865 5G   \\
 \hline

  \rule{0pt}{2ex}BLU C5L Max  & 11 & 2021 & Unisoc SC9832   \\
 \hline

   \rule{0pt}{2ex} \multirow{ 2}{*}{TCL 30}  &  \multirow{ 2}{*}{12} &  \multirow{ 2}{*}{2022} & Mediatek MT6765V/CB Helio G37    \\
 \hline

    \rule{0pt}{2ex} \multirow{ 2}{*}{Samsung Galaxy S8+} &  \multirow{ 2}{*}{9} &  \multirow{ 2}{*}{2018} & Qualcomm MSM8998 Snapdragon 835  \\
 \hline

    \rule{0pt}{2ex}Motorola Moto G Play   & 12 &  2022 &  Mediatek MT6765 Helio G37   \\
 \hline
\end{tabular}
\label{tab:ld}
\end{table}
\vspace{-0.5cm}
\end{center}

\noindent \textbf{Implication.} NCC and NH are critical parameters that are used to establish AS security and derive $K_{gnb}$. These parameters are used during the \packet{RRC\_Reestablish} procedure to re-establish the RRC connection. This procedure is particularly important during handover. The expectation is that these session key creation parameters (NCC and NH) would be fresh and unused to create diverse keys (precisely described in $T_1$). This is an important assumption ensuring forward and future secrecy guarantees of the keys. Forward and future secrecy ensures that the protocol defends the past and future sessions even if the current session is compromised~\cite{bluffs}. These parameters are essentially used as nonces to ensure the diversification of the keys. However, if the fresh and unused NCC/NH value usage is not mandated, then the forward and future secrecy guarantees can be broken. Furthermore, as the \packet{RRC\_Reestablish} message (containing these parameters) is unencrypted, the attacker can easily detect the sessions where the same NCC/NH values are used for key derivation. 

\noindent \textbf{Investigation.} 
We found that none of the open-source implementations properly check these parameters and just accept if they differ from the previously accepted ones.




\begin{center}

\def\arraystretch{1}
\begin{table*}[ht]
	\centering
    \caption{COTS UE behavior based on the inconsistencies found by \system{}. \textcolor{red}{Red} marks the Exploitable cases. \neutral{\unverified{}} denotes cases where the result could not be verified.}
    \captionsetup{font={small}}
	\renewcommand{\arraystretch}{1}
	\fontsize{7}{7}\selectfont
\begin{tabular}{P{0.1\linewidth}||P{0.07\linewidth}P{0.07\linewidth}P{0.07\linewidth}P{0.07\linewidth}P{0.07\linewidth} P{0.06\linewidth}  P{0.06\linewidth}  P{0.06\linewidth} P{0.06\linewidth} } 
\rule{0pt}{2ex} Device  & \multicolumn{2}{c|}{F1} & \multicolumn{1}{|c|}{F2} &  \multicolumn{2}{c|}{F3} & \multicolumn{3}{c|}{F5} & F6 \\ [0.5ex] 
 \hline\hline
 \rule{0pt}{2ex}  & Plain-text \packet{auth} \packet{request} accepted & Plain-text \packet{identity} \packet{request} accepted & Plain-text \packet{detach} \packet{request} accepted & Integrity-failed message accepted & Causes connection drop & \packet{attach} \packet{reject} clears context & \packet{service} \packet{reject} clears context & \packet{tau} \packet{reject} clears context & \packet{tau} \& \packet{detach} collision\\
 \hline

 \rule{0pt}{2ex}  \multirow{ 2}{*}{Google Pixel 7a} & \dno\multirow{ 2}{*}{\unsafe{\cmark}} &  \multirow{ 2}{*}{\xmark} & \multirow{ 2}{*}{\xmark}  & \multirow{ 2}{*}{\xmark} & \dno\multirow{ 2}{*}{\unsafe{\cmark}} & \multirow{ 2}{*}{\cmark}  & \dno\multirow{ 2}{*}{\unsafe{\xmark}} & \dno\multirow{ 2}{*}{\unsafe{\xmark}} & \packet{TAU} progressed\\
 \hline

 \rule{0pt}{2ex}Samsung Galaxy S20 FE & \multirow{ 2}{*}{\xmark} &  \multirow{ 2}{*}{\xmark} & \multirow{ 2}{*}{\xmark}  & \multirow{ 2}{*}{\xmark} & \dno\multirow{ 2}{*}{\unsafe{\cmark}} & \multirow{ 2}{*}{\cmark} &  \dno\multirow{ 2}{*}{\unsafe{\xmark}} & \dno\multirow{ 2}{*}{\unsafe{\xmark}} & \packet{TAU} progressed\\
 \hline

 \rule{0pt}{2ex} HTC One E9 Plus & \dno\multirow{ 2}{*}{\unsafe{\cmark}} &  \dno\multirow{ 2}{*}{\unsafe{\cmark}} & \multirow{ 2}{*}{\xmark}  & \multirow{ 2}{*}{\xmark} & \dno\multirow{ 2}{*}{\unsafe{\cmark}} & \multirow{ 2}{*}{\cmark}  & \dno\multirow{ 2}{*}{\unsafe{\xmark}} & \dno\multirow{ 2}{*}{\unsafe{\xmark}} & \packet{TAU} progressed\\
 \hline

 \rule{0pt}{2ex}  \multirow{ 2}{*}{Huawei Y5} & \dno\multirow{ 2}{*}{\unsafe{\cmark}} &  \dno\multirow{ 2}{*}{\unsafe{\cmark}} & \multirow{ 2}{*}{\xmark}  & \multirow{ 2}{*}{\xmark} & \dno\multirow{ 2}{*}{\unsafe{\cmark}} & \multirow{ 2}{*}{\cmark}  & \dno\multirow{ 2}{*}{\unsafe{\xmark}} & \dno\multirow{ 2}{*}{\unsafe{\xmark}} & \packet{TAU} progressed\\
 \hline

 \rule{0pt}{2ex}  \multirow{ 2}{*}{Xiaomi 11 Lite} & \multirow{ 2}{*}{\xmark} &  \multirow{ 2}{*}{\xmark} & \multirow{ 2}{*}{\xmark}  & \multirow{ 2}{*}{\xmark} & \dno\multirow{ 2}{*}{\unsafe{\cmark}} & \multirow{ 2}{*}{\cmark} &  \dno\multirow{ 2}{*}{\unsafe{\xmark}} & \dno\multirow{ 2}{*}{\unsafe{\xmark}} & \packet{TAU} progressed\\
 \hline

 \rule{0pt}{2ex} Motorola Edge 30 Pro & \multirow{ 2}{*}{\xmark} &  \multirow{ 2}{*}{\xmark} & \multirow{ 2}{*}{\xmark}  & \multirow{ 2}{*}{\xmark} & \dno\multirow{ 2}{*}{\unsafe{\cmark}} & \multirow{ 2}{*}{\cmark}  & \dno\multirow{ 2}{*}{\unsafe{\xmark}} & \dno\multirow{ 2}{*}{\unsafe{\xmark}} & \packet{TAU} progressed\\
 \hline

  \rule{0pt}{2ex}  \multirow{ 2}{*}{OnePlus 9 Pro} & \multirow{ 2}{*}{\xmark} &  \multirow{ 2}{*}{\xmark} & \multirow{ 2}{*}{\xmark}  & \multirow{ 2}{*}{\xmark} & \dno\multirow{ 2}{*}{\unsafe{\cmark}} & \multirow{ 2}{*}{\cmark} & \dno\multirow{ 2}{*}{\unsafe{\xmark}} & \dno\multirow{ 2}{*}{\unsafe{\xmark}} & \packet{TAU} progressed\\
 \hline

  \rule{0pt}{2ex} Huawei Honor 8X & \multirow{ 2}{*}{\xmark} &  \multirow{ 2}{*}{\xmark} & \multirow{ 2}{*}{\xmark}  & \multirow{ 2}{*}{\xmark} & \multirow{ 2}{*}{\xmark} & \multirow{ 2}{*}{\cmark} & \dno\multirow{ 2}{*}{\unsafe{\xmark}} & \dno\multirow{ 2}{*}{\unsafe{\xmark}} & \packet{TAU} progressed\\
 \hline

  \rule{0pt}{2ex} Apple iPhone 12 Pro & \multirow{ 2}{*}{\xmark} &  \multirow{ 2}{*}{\xmark} & \multirow{ 2}{*}{\xmark}  & \multirow{ 2}{*}{\xmark} & \dno\multirow{ 2}{*}{\unsafe{\cmark}} & \multirow{ 2}{*}{\cmark} & \multirow{ 2}{*}{\cmark} & \dunverified\multirow{ 2}{*}{\neutral{\unverified{}}} & \dunverified\multirow{ 2}{*}{\neutral{\unverified{}}}\\
\hline 

   \rule{0pt}{2ex}  \multirow{ 2}{*}{Google Pixel 3a} & \multirow{ 2}{*}{\xmark} &  \multirow{ 2}{*}{\xmark} & \multirow{ 2}{*}{\xmark}  & \multirow{ 2}{*}{\xmark} & \dno\multirow{ 2}{*}{\unsafe{\cmark}} & \multirow{ 2}{*}{\cmark} &  \dno\multirow{ 2}{*}{\unsafe{\xmark}} & \dno\multirow{ 2}{*}{\unsafe{\xmark}} & \packet{TAU} progressed\\
 \hline

   \rule{0pt}{2ex} Samsung Galaxy A04 & \multirow{ 2}{*}{\xmark} &  \multirow{ 2}{*}{\xmark} & \multirow{ 2}{*}{\xmark}  & \multirow{ 2}{*}{\xmark} & \dno\multirow{ 2}{*}{\unsafe{\cmark}} & \multirow{ 2}{*}{\cmark} &  \dno\multirow{ 2}{*}{\unsafe{\xmark}} & \dno\multirow{ 2}{*}{\unsafe{\xmark}} & \packet{TAU} progressed\\
 \hline

  \rule{0pt}{2ex}  \multirow{ 2}{*}{LG Velvet 5G} & \dno\multirow{ 2}{*}{\unsafe{\cmark}} &  \dno\multirow{ 2}{*}{\unsafe{\cmark}} & \multirow{ 2}{*}{\xmark}  & \multirow{ 2}{*}{\xmark} & \dno\multirow{ 2}{*}{\unsafe{\cmark}} & \multirow{ 2}{*}{\cmark}  & \dno\multirow{ 2}{*}{\unsafe{\xmark}} & \dno\multirow{ 2}{*}{\unsafe{\xmark}} & \packet{TAU} progressed\\
 \hline

   \rule{0pt}{2ex}  \multirow{ 2}{*}{OnePlus 8T} & \multirow{ 2}{*}{\xmark} &  \multirow{ 2}{*}{\xmark} & \multirow{ 2}{*}{\xmark}  & \multirow{ 2}{*}{\xmark} & \dno\multirow{ 2}{*}{\unsafe{\cmark}} & \multirow{ 2}{*}{\cmark} & \dno\multirow{ 2}{*}{\unsafe{\xmark}} & \dno\multirow{ 2}{*}{\unsafe{\xmark}} & \packet{TAU} progressed\\
 \hline

   \rule{0pt}{2ex}  \multirow{ 2}{*}{BLU C5L Max} & \multirow{ 2}{*}{\xmark} &  \multirow{ 2}{*}{\xmark} & \multirow{ 2}{*}{\xmark}  & \multirow{ 2}{*}{\xmark} & \dno\multirow{ 2}{*}{\unsafe{\cmark}} & \multirow{ 2}{*}{\cmark} & \dno\multirow{ 2}{*}{\unsafe{\xmark}} & \dno\multirow{ 2}{*}{\unsafe{\xmark}} & \packet{TAU} progressed\\
 \hline

    \rule{0pt}{2ex}  \multirow{ 2}{*}{TCL 30} & \multirow{ 2}{*}{\xmark} &  \multirow{ 2}{*}{\xmark} & \multirow{ 2}{*}{\xmark}  & \multirow{ 2}{*}{\xmark} & \dno\multirow{ 2}{*}{\unsafe{\cmark}} & \multirow{ 2}{*}{\cmark} & \dno\multirow{ 2}{*}{\unsafe{\xmark}} & \dno\multirow{ 2}{*}{\unsafe{\xmark}} & \packet{TAU} progressed\\
 \hline

       \rule{0pt}{2ex} Samsung Galaxy S8+ & \multirow{ 2}{*}{\xmark} &  \multirow{ 2}{*}{\xmark} & \multirow{ 2}{*}{\xmark}  & \multirow{ 2}{*}{\xmark} & \dno\multirow{ 2}{*}{\unsafe{\cmark}} & \multirow{ 2}{*}{\cmark} &  \dno\multirow{ 2}{*}{\unsafe{\xmark}} & \dno\multirow{ 2}{*}{\unsafe{\xmark}} & \packet{TAU} progressed\\
\hline

    \rule{0pt}{2ex} Motorola Moto G Play & \multirow{ 2}{*}{\xmark} &  \multirow{ 2}{*}{\xmark} & \multirow{ 2}{*}{\xmark}  & \multirow{ 2}{*}{\xmark} & \dno\multirow{ 2}{*}{\unsafe{\cmark}} & \multirow{ 2}{*}{\cmark} &  \dno\multirow{ 2}{*}{\unsafe{\xmark}} & \dno\multirow{ 2}{*}{\unsafe{\xmark}} & \packet{TAU} progressed\\
 \hline

 \hline

\end{tabular}
\label{tab:uetest}
\end{table*}
\end{center}

\subsubsection{Privacy}

\stepcounter{fdcounter}
\noindent\findingsty{\thefdcounter. }\textit{GUTI deletion.}
On many occasions, when a reject message is received from the network, the UE updates its EPS/5GS status, clears the context, and subsequently moves to a deregistered state. We observe that in many cases, a rejection cause $\mathcal{T}_\mathcal{C}$ in a reject message $\mathcal{R}_\mathcal{M}$ would suggest the UE to clear the context and move to deregistered state while the same cause $\mathcal{T}_\mathcal{C}$ would keep the UE in registered state without clearing the security context (shown in Figure~\ref{lt:5.1}). For example, 5GM cause \#13 received through 
\packet{registration\_reject} suggests the UE to delete GUTI, TAI, ngKSI while the same cause received through \packet{tau\_reject} or \packet{service\_reject} message would keep the UE in a registered state without deleting the context. A cross-examination may suggest that since the messages are different, the same cause may trigger different behavior. However, other cause values such as \#12 (tracking area not allowed) trigger similar behavior and state transition for different reject messages. Thus, it concretely verifies the conflicting suggestions about security context. 

\begin{figure}[htbp]
\vspace{-0.5cm}
    \centering
   \noindent\begin{mdframed}[font=\scriptsize]
    $T_1$: \#13 (Roaming not allowed in this tracking area) The UE shall set the 5GS update status to 5U3 ROAMING NOT ALLOWED (and shall store it according to subclause5.1.3.2.2) and shall \textbf{delete 5G-GUTI, last visited registered TAI, TAI list and ngKSI}
    \\
    \noindent $T_2$: \#13 (Roaming not allowed in this tracking area) The UE shall set the 5GS update status to 5U3 ROAMING NOT ALLOWED (and shall store it according to subclause5.1.3.2.2) and shall delete the list of equivalent PLMNs (if available).
    \end{mdframed}
    \caption{GUTI, TAI, eKSI deletion. $T_1$ is from section 5.5.1.2.5 and $T_2$ is from section 5.5.1.3.5 of TS 24.501.}
    \label{lt:5.1}
    \vspace{-0.5cm}
\end{figure}

\noindent \textbf{Inconsistency to exploit.}
GUTI (Globally Unique Temporary ID) is a kind of temporary ID used to identify UEs. Each UE has a couple of different kinds of unique IDs, like IMSI, IMEI, etc. These sorts of temporary identifiers, like GUTI, are used to prevent attackers from tracking users. However, if these identifiers are not changed or reset, this can cause several privacy issues. An adversary can utilize old GUTI to track a UE through a linkability attack, violating privacy~\cite{Hong2018GUTIRD}.

 \noindent \textbf{Investigation.} 
 In commercial UEs, we found a total of 16 devices not properly deleting GUTI for these reject messages (Table~\ref{tab:uetest}). One recent UE (Apple iPhone 12 Pro), however, deletes the security context with GUTI after receiving \packet{service\_reject}.

\subsubsection{Denial of Service}

\stepcounter{fdcounter}
\noindent\findingsty{\thefdcounter}. \textit{TAU and detach collision.}
We found a PoS of conflicting directions when TAU and detach procedure collide (shown in Figure~\ref{lt:2.1}).
The first one states that the \packet{tracking\_area\_update} procedure shall be aborted and the \packet{detach} procedure shall be progressed. On the other hand, the second segment states that the \packet{detach} procedure shall be aborted and re-initiated while the \packet{tracking\_area\_update} procedure is fully performed.


\begin{figure}[htbp]
    \centering
    \vspace{-0.5cm}
\noindent\begin{mdframed}[font=\scriptsize]
$T_1$: Tracking area updating and detach procedure collision \\
EPS detach containing detach type "re-attach required"
or "re-attach not required": If the UE receives a DETACH
REQUEST message before the tracking area updating
procedure has been completed, the \textbf{tracking area updating procedure shall be aborted} and the \textbf{detach procedure
shall be progressed}.
\\
\noindent $T_2$: If a cell change
into a new tracking area that is not in the stored TAI
list occurs before the UE initiated detach procedure is
completed, the UE proceeds as follows: 1) If the detach
procedure was initiated for reasons other than removal
of the USIM or the UE is to be switched off, the \textbf{detach
procedure shall be aborted} and re-initiated \textbf{after success-
fully performing a tracking area updating procedure}.
\end{mdframed}
\vspace{-0.5cm}
    \caption{TAU and Detach Procedure precedence conflicts. $T_1$ is from section 5.5.3.2.6 and $T_2$ is from section 5.5.2.2.4
of TS 24.301.
    }
    \label{lt:2.1}
\end{figure}

\noindent \textbf{Inconsistency to exploit.} An attacker can achieve downgrade/denial-of-service by injecting \packet{detach\_request} through fake-base-stations or signal-injection attacks~\cite{Hussain2018LTEInspectorAS}. In this case, an implementation that aims to implement both scenarios can exacerbate the situation by creating a deadlock state.
%
As the statements are mutually exclusive either way, an implementation violates the specification. In our investigation of commercial implementation, we found all the UEs progress with the \packet{tracking\_area\_update} procedure when such collision occurs (Table~\ref{tab:uetest}).

\subsubsection{Interoperability}
For this class of inconsistencies, we could not devise an attack scenario. However, these inconsistencies range from specifying different sub-state transitions for the same condition to missing timer expiration directives. The potential impact of these inconsistencies can cause interoperability issues between different implementations, i.e., one implementation following one directive and the other one following a different directive. For brevity, one of them is discussed here; we discuss the rest in~\cite{cellularlint}.

\stepcounter{fdcounter}
\looseness = -1
\noindent\findingsty{\thefdcounter.} \textit{Sub-state transition confusion.}
When \packet{attach\_reject} message is received with the emm cause \#14, the 4G specification has differing state transition descriptions (shown in Figure~\ref{lt:1.1}). This certainly can cause confusion in implementation design.

\noindent \textbf{Investigation.} While looking further into open-source implementations, we found a very interesting scenario regarding this. In srsRAN, we see that the developers tried to implement both of them. We show the code snippet of srsRAN in Figure~\ref{lt:1.2}. Here, we can see that both of the conditional statements contain the same cause (lines 1 and 6) but have different sub-state transitions (lines 3 and 7). Of course, UE will go to the \textit{EMM\_DEREGISTERED.PLMN\_SEARCH} sub-state due to the obscurity of the first "if". Such conflicts can potentially cause interoperability issues, leaving a possibility for further synchronization problems.  


\lstset{basicstyle=\footnotesize\ttfamily,
columns=fullflexible, 
breaklines=true
,frame=single,
backgroundcolor=\color{backcolour},   
commentstyle=\color{codegreen},
keywordstyle=\color{codepurple},
numberstyle=\tiny\color{codegray},
stringstyle=\color{blue},
escapeinside={<@}{@>}}

\begin{figure}
    \centering
    \begin{lstlisting}[linewidth=\columnwidth,language=C++]
 1   if (...||  attach_rej.emm_cause == <@\textcolor{blue}{LIBLTE\_MME\_EMM\_CAUSE\_EPS\_SERVICES\_NOT\_}@> <@\textcolor{blue}{ALLOWED\_IN\_THIS\_PLMN}@> ||...){
 2       <@\ \quad @>attach_attempt_counter = 0;
 3       <@\ \quad@>enter_emm_deregistered(emm_state_t::deregistered_substate_t::<@\textcolor{red}{attempting\_to\_attach}@>);
 4   }
 5   // TODO: handle other relevant reject causes
 6   if (...|| attach_rej.emm_cause == <@\textcolor{blue}{LIBLTE\_MME\_EMM\_CAUSE\_EPS\_SERVICES\_NOT\_}@> <@\textcolor{blue}{ALLOWED\_IN\_THIS\_PLMN}@> ||...) {
 7       <@\ \quad@>enter_emm_deregistered(emm_state_t::deregistered_substate_t::<@\textcolor{red}{plmn\_search}@>);
  }
    \end{lstlisting}
    \caption{srsRAN implementation of emm cause \#14} 
    \label{lt:1.2}
\end{figure}




\stepcounter{fdcounter}
\noindent\findingsty{\thefdcounter.} \textit{Registration counter reset.}
When a UE receives a \packet{registration\_reject} message with several cause such as \#11 (PLMN not allowed) it resets the registration attempt counter. For the same cause value received through \packet{service\_reject} message, the UE should reset the associated service request attempt counter. However, this directive is missing in the specification, although in the same 5G NAS specification, the general guideline for \packet{service\_reject} with a list of causes clearly suggests resetting the counter. Failure to clarify this in corresponding cause descriptions has an impact on interoperability as some implementations take the general guideline into account while others focus on specific scenarios.

\stepcounter{fdcounter}
\noindent\findingsty{\thefdcounter. }\textit{Timer expiry as a precondition.}
When a \packet{registration\_reject} is received in case of 5G Standalone Non-Public Network (SNPN), the "list of subscriber data" is considered invalid until switch off or entry update or timer T3245 expiration. However, the \packet{authentication\_reject} under same condition does not consider the time T3245 expiry for the list validity. This is another conundrum causing interoperability issues.

\stepcounter{fdcounter}
\noindent\findingsty{\thefdcounter. }\textit{Unutilized service attempt counter. }
We observe that the service request attempt counter is less utilized compared to the registration attempt or tracking area update attempt counter for similar events. When a registration attempt is rejected with proper cause value, the counter is reset, but for many \packet{service\_reject} the service attempt counter is completely ignored in the description. This may potentially cause interoperability among different vendor implementations.

\stepcounter{fdcounter}
\noindent\findingsty{\thefdcounter. }\textit{PDCP counter set. }
We observed that for the protection of RRC messages while transitioning away from \textit{RRC\_INACTIVE} to \textit{RRC\_CONNECTED}, UE derives a set of keys such as $K_{RRCint}$, $K_{RRCenc}$, and so on. However, despite having similar actions of key deriving for connection resume in \textit{CM-IDLE}, the UE additionally resets the PDCP count to 0 and activates new AS keys in PDCP layer. This final action is totally skipped in the first case (Fig~\ref{lt:pdcp}).

\noindent \textbf{Impact.} This subtle miss can have a significant de-synchronization impact if the COTS implementation considers one of them ignoring the other.

\begin{figure}[ht]
    \centering
\noindent\begin{mdframed}[font=\footnotesize]
\noindent$T_1$: For protection of all RRC messages except \packet{RRCReject} message following the sent \packet{RRCResumeRequest} message, the UE shall derive a ${K_{NG-RAN}}^*$ using the target PCI, target ARFCN-DL/EARFCN-DL and the $K_{gNB}/NH$ based on either a horizontal key derivation or a vertical key derivation as defined in clause 6.9.2.1.1 and Annex A.11/Annex A.12. The UE shall further derive $K_{RRCint}$, $K_{RRCenc}$, $K_{UPenc}$ (optionally), and $K_{UPint}$ (optionally) from the newly derived ${K_{NG-RAN}}^*$.
\\

\noindent$T_2$: For protection of all RRC messages except RRC Reject message following the sent RRC Resume Request message, the UE shall derive a ${K_{NG-RAN}}^*$ using the target PCI, target EARFCN-DL and the $K_{gNB}/NH$ based on either a horizontal key derivation or a vertical key derivation as defined in clause 6.9.2.1.1 and Annex A.12. The UE shall further derive $K_{RRCint}$, $K_{RRCenc}$, $K_{UPenc}$ (optionally), and $K_{UPint}$ (optionally) from the newly derived ${K_{NG-RAN}}^*$. \textbf{Then the UE resets all PDCP COUNTs to 0 and activates the new AS keys in PDCP layer}.
\end{mdframed}
    \caption{PDCP counter reset ignored. $T_1$ is from section 6.8.2.1.3 of TS 33.501 and $T_2$ is from section 6.16.2.3 of TS 33.501  
    }
    \label{lt:pdcp}
\end{figure}




\subsection{RQ4. Comparison of other methodologies}
\begin{center}
\begin{table}[ht]
\centering
\fontsize{6}{6}\selectfont
\caption{Comparison with existing approaches. }
\captionsetup{font={small}}
\caption*{\filledrect{red!90}: Did not discover\ \qquad \filledrect{cyan!90}: Closely related issue discovered\qquad\quad \filledrect{orange!90}: Only context is similar\quad\filledrect{mygreen!90}: Discovered}
\begin{tabular}{p{0.08\linewidth}|p{0.12\linewidth}|p{0.11\linewidth}|p{0.11\linewidth} |p{0.14\linewidth}|p{0.13\linewidth}} 

 Findings & DIKEUE\cite{10.1145/3460120.3485388} & DoLTEst\cite{277258}  & ATOMIC\cite{9519388} & Instructions Unclear\cite{291203}  & \system{}\\ [0.5ex] 
 \hline\hline

 1 & \no & \no & \no & \no & \yes \\
 \hline
 2 & \yes & \related & \no & \yes & \yes\\
  \hline
 3 & \yes & \yes & \related & \related & \yes \\ 
 \hline
 4 & \yes & \yes & \yes & \related & \yes \\
  \hline
 5 & \related & \related & \yes & \no & \yes \\ 
  \hline
 6 & \no & \no & \discussion & \no & \yes \\
 \hline
 7 & \no & \no & \discussion & \no & \yes\\ 
 \hline
 8~\cite{cellularlint} & \no & \discussion & \related & \discussion & \yes \\ 
  \hline
 9~\cite{cellularlint} & \no & \no & \no & \discussion & \yes \\ 
 \hline
 10~\cite{cellularlint} & \discussion & \discussion & \related & \no & \yes\\ 
 \hline
 11~\cite{cellularlint} & \no & \related & \no & \no & \yes \\
\end{tabular}

\label{tab:compare methods}
\end{table}
\end{center}
\vspace{-0.5cm}

Although to the best of our knowledge \system{} is the first approach to detect inconsistencies from cellular specifications, we compare our methodology with the existing approaches based on the issues discovered. Table~\ref{tab:compare methods} shows comparison results based on the subset of findings presented in the paper. In terms of methodologies we compare with DIKEUE and DoLTEst that are implementation testing frameworks and, therefore, are not directly comparable to our methodology. The closest methods that deal with specification descriptions are~\cite{9519388} and~\cite{291203}. We compare with respect to four characteristics: (1) if the issue was discovered, (2) if the context aligned to the issue has been discussed, (3) if a related root cause has been discussed, and (4) if the issue has not been discussed at all.

%% file: sections/related_works.tex
\section{Related Works}\label{sec:related works}
We discuss the related works in three categories:







\noindent \textbf{NLP efforts. }
Sequence classification tasks, especially supervised textual entailment, have largely been used recently in various domains and applications~\cite{hu-etal-2020-ocnli,9003090,snli:emnlp2015,9519388,sadat2022scinli}. In contrast to 
those techniques, Andow et al.~\cite{236198,247632} use Named Entity Recognition to leverage a rule-based approach for detecting privacy leaks in privacy policies of Android apps. The method, however, considers domain-specific assumptions for ontology structuring and is not generalizable to cellular protocol documents.

\looseness = -1
\noindent \textbf{NLP for cellular network. }
In recent days, NLP research in cellular networks has been gaining popularity due to the enormous information and the emergence of large language models. For instance, methodologies have been proposed to automatically analyze specification, RFC, and Change Request documents in various problem setting~\cite{ishtiaq2023hermes,9519388,279972,287360,9833673,karim-EtAl:2023:findings}. Pacheco et al.~\cite{9833673} designed an approach that automatically builds state machines from RFC protocol documents using a data-driven zero-shot approach and synthesizes attacks based on that. Ishtiaq et al.~\cite{ishtiaq2023hermes} take a similar approach to synthesize FSM from 4G and 5G specifications. Chen et al.~\cite{9519388} discover Hazard Indicators (HI) from 4G specifications and verify implementations against them. Briefly, they input Risky Operation Description (ROD) and threat models in their framework and run a textual entailment model to find the HIs from which the test cases are generated. While they generate test cases from individual text descriptions, our method does a differential analysis on pairs of descriptions to generate test cases.  A conformance testing approach has also been proposed based on NLP and causal relation extraction~\cite{287360}. However, these methods do not primarily focus on inconsistencies in standards.

\looseness = -1
\noindent \textbf{Vulnerability detection in cellular networks:}
Over the years a lot of work has focused on finding issues in various generations of cellular network design and implementation 
using formal verification~\cite{10.1145/3319535.3354263,DBLP:conf/ndss/CremersD19}, dynamic analysis~\cite{8835363},  differential testing~\cite{10.1145/3460120.3485388}, fuzzing, etc.  Klischies et al.~\cite{291203} consider underspecifications from protocol standards that are closely aligned with our goal. Baseband firmware analysis has also been effective in finding implementation issues~\cite{179505,10.1145/3395351.3399360,Kim2021BaseSpecCA}. However, these methods do not relate the root causes of implementation issues to the protocol standards.

%% file: sections/discussion.tex
\section{Discussion}\label{sec:dis}

\textbf{Responsible Disclosure.}
To enhance the standards of cellular networks, we have conscientiously disclosed all our \numissuesclean{} findings to both the GSMA and the affected User Equipment (UE) vendors. The GSMA's Coordinated Vulnerability Disclosure (CVD) panel acknowledged that the specifications are not flawless and may contain inconsistencies within the current versions. Additionally, it is not in the scope of their mandate to act on the alleged errors in open-source mobile network implementations. However, they also believe that inconsistencies do not lead to security and privacy issues, and developers can accurately implement the protocol by referring to additional specifications for more detailed information.
For instance, to address the confusion found in finding 11, developers should consult both the Radio Resource Control (RRC) specification (TS 38.331) and the Packet Data Convergence Protocol (PDCP) specification (TS 38.321). The panel also pointed out that varying UE implementations might handle situations differently when the specifications do not prescribe a specific behavior. 

Nonetheless, our empirical results derived from testing open-source implementations and various commercial devices indicate otherwise. We clearly observed that different implementations stem from ambiguous and inconsistent descriptions within the specifications and can lead to severe security and privacy issues.
This inconsistency can potentially be mitigated through a meticulous revision of the standards, ensuring clearer guidance and uniformity across different implementations.

\noindent \textbf{Manual efforts and limitations in \system{}. } Although \system{} can discover conflicts under weakly supervised settings, there is still some manual labor involved. During the active learning phase, the annotators have to cross-examine the predictions from the model and refine them accordingly. This required approximately 16 human hours from domain experts. Furthermore, one of the critical limitations of \system{} is that there are certain inconsistent PoS that, when taken into a broader specification context, are no longer inconsistent. This issue, with a larger context, requires manual efforts to parse through the results of \system{} and filter out. This, in some sense, is the limitation of current NLP models that are unable to take into account this large context of cellular network specifications. To provide an example, in some cases, the context is defined 3-4 pages before the actual segment. It took about 6 human hours to detect the false positive from the final results.  


\noindent \textbf{Inconsistency-to-attack scenario.} The aim of \system{} is to find inconsistencies in the 4G and 5G specifications. As mentioned in \sectionref{subsec:impact}, the design from inconsistent PoS to concrete attacks is a manual effort. We look into the POSs to uncover whether these design choices can lead to an attack. We want to emphasize that not all inconsistencies can lead to exploitable attacks. Similarly, as this part of the findings is manually designed, there can be other inconsistencies that lead to exploitable attacks but are not discussed in detail in the paper. Therefore, we open-source the list of major inconsistencies~\cite{cellularlint}.

\noindent \textbf{Implementation vs. specification issues.} In terms of security issues in the 4G and 5G, they can be characterized in either one of the two categories: issues stemming from the specification design or 
from the implementation design. 
In this paper, we focus on the first category. For instance, the issues in findings 1 and 3 can be categorized as implementation issues where the UE accepts plain-text or integrity-failed messages. However, we focus on the specification consistency. The aim of our work is to clear scenarios in the standards that can cause inconsistencies in the implementation. We believe our work would inspire further specification-based textual analysis and ultimately realize the goal of machine-readable specifications.

\noindent \textbf{Scalability and relevance to further security research.} \system{} is designed in a fashion that can be almost readily used for other cellular specifications and, with some training effort, can be utilized for other communication (non-cellular) protocols. For other cellular specifications, one can use the pre-trained+SNLI-fine-tuned state of~\sysmodel{}. Only the last fine-tuning step with a low amount of data would be required. For completely different communication protocols, one can utilize the methodology of \system{} to pre-train and fine-tune the models. Pairing and filtration, PoS selection, and optimization techniques would still follow the same methods. Furthermore, the impact of \system{} is not limited to cellular network security, it can be applied to enhance the security of other communication protocols as well.

\noindent \textbf{Impact of tables, figures and codes.} While preprocessing, we removed tabular data, figures, and code snippets. These artifacts can potentially add more information to decide which of the text pairs are more secure. For example, \system{} captured an inconsistent description regarding EAP-TLS authentication procedure, which we had to verify as a false positive manually. However, figure B.2.1.1-1 under Annex B.2 in TS 33.501 describes a timing diagram of that authentication procedure more clearly. Considering this could potentially help \system{} capture the false positive in initial stage of analysis.
Moreover, developers often focus on the finite state machines presented through the figures. Thus, learning from these artifacts alongside texts could strengthen the learning process, and we leave it as future work.

%% file: sections/conclusion.tex
\section{Conclusion and future work}\label{sec:conclusion}

In the paper, we propose \system{} to detect inconsistencies in cellular protocol specifications in a scalable fashion. \system{} uncovers~\numissuesclean{} inconsistencies with $82.67\%$ accuracy in the NAS and security specifications. After verification of these inconsistencies on open-source implementations and commercial devices, we confirm that they indeed have a substantial impact on design decisions, potentially leading to concerns related to privacy, integrity, availability, and interoperability. 

\noindent \textbf{Future work.} In the future, we will work on improving \system{} to include further context and improve the accuracy. Furthermore, we will aim to detect underspecifications with NLP techniques.

%% file: sections/appendix.tex
\section{Appendix}\label{sec:app}

\subsection{Testbed \& devices details}\label{subsec:ld}
\noindent \textbf{Testbed Preparation.} \label{subsubsec:tp}
To test the finalized inconsistent-tagged PoS in open-source and Commercial-Of-The-Shelf (COTS) UE implementations, we follow different setups for 4G and 5G. For 4G, we use USRP B210 with srsRAN eNB and core on one side and either (1) another USRP B210 with srsRAN UE, or (2) COTS UE with LTE capabilities on the other side. To run for 5G test cases, we use USRP B210 with srsRAN gNB and Open5GS core network on one side and either (1) another USRP B210 with srsRAN UE, or (2) COTS UE with 5G NR capabilities on the other side. In both scenarios, we use Leo Bodnar GPSDO to generate unjittered 10 Mhz frequency, which helps a more stable connection--often essential for 5G setups. 

To establish a connection and send specific packets in LTE, we configure custom messages over srsRAN implementations. For 5G testing, we follow the open-source implementation demonstrated by~\cite{bitsikas23UEframework} to tailor the srsRAN and open5GS codebase in order to accommodate custom message passing and parsing based on our needs.

Table~\ref{tab:ld} details the information of the devices tested by~\system{}

\subsection{Network Control Plane Procedures}\label{subsec:control}
\begin{figure}[ht]
    \centering
    \includegraphics[width= 1.1\linewidth, trim = 0cm 14cm 0cm 0cm, clip]{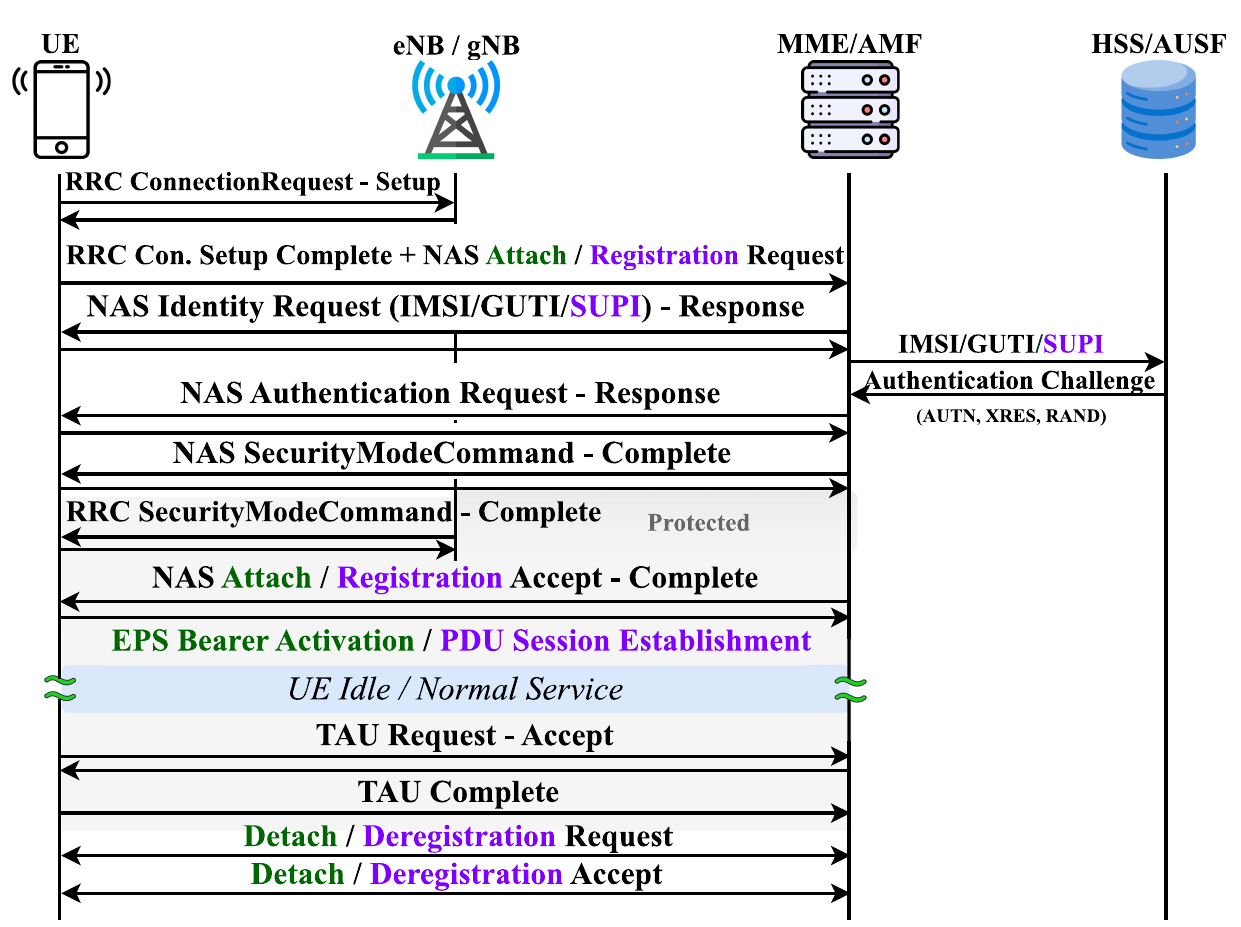}
    \caption{Cellular Network Control Plane Procedures. Procedures unique to 4G and 5G are shown in \textcolor{greener}{\textbf{green}} and \textcolor{myviolet}{\textbf{purple}} respectively.
    }
    \label{fig:back:4G}
\end{figure}
The control plane procedures are responsible for authentication, maintenance, and management the control and signaling aspects of communication between network elements and UE. 
Fig.~\ref{fig:back:4G} represents the control plane procedures for cellular networks. For the sake of brevity, 4G and 5G authentication and handover are discussed here in a consolidated manner. 

When a UE wants to connect to a cellular network (e.g., after powering on or moving into a new coverage area), at first, the UE and the nearest eNB/gNB completes the Radio Resource Control (RRC) setup procedures. Next, the UE sends an \packet{attach\_request} to that base station. 
The eNB forwards the \packet{attach\_request} to the MME. The MME, in turn, initiates the authentication process by sending an \packet{identity\_request} message to the UE--asking for the IMSI. If the UE responds with its IMSI in the \packet{identity\_response} message, the MME forwards it to the HSS. The HSS replies with an authentication challenge which contains a random number (RAND) and Authentication Token (AUTN), which is forwarded to the UE using the \packet{authentication\_request} message. The UE uses this RAND along with its secret key $(K_i)$ stored in the USIM to calculate the Expected Authentication Response (XRES) using an Authentication and Key Agreement (AKA) algorithm.

The UE uses the AUTN to authenticate the core network. Next, The UE sends the calculated XRES to the MME in an \packet{authentication\_response} message. MME validates the received XRES and derives the secret keys for secure communication. Next, the \packet{security\_mode\_command} and  \packet{security\_mode\_complete} messages are used to share the secret keys and the connection is completed through \packet{attach\_accept} and \packet{attach\_complete} messages. At any point, the UE or the MME can initiate a disconnection through a \packet{detach\_request} message for which the other party can optionally respond with a \packet{detach\_accept} message depending on the cause value associated with detach.

For 5G, the \packet{attach\_request} has been replaced by \packet{registration\_request} which is forwarded by the gNB to the AMF. Here, the UE sends a Subscription Permanent Identifier (SUPI) instead of IMSI which has enhanced privacy capability. The UE may send a Subscription Concealed Identifier (SUCI) instead of plain SUPI when authenticating for the first time. Similar to 4G, After authenticating each other and exchanging \packet{security\_mode\_command} and \packet{security\_mode\_complete} the registration is completed. The detach procedures of 4G is replaced by deregistration procedures in 5G.


When a UE moves from one location to another geographically, it sends a Tracking Area Update request (\packet{tau\_request}) to the network. This request usually contains the new Tracking Area Identity (TAI). The network responds with a \packet{tau\_accept} and the UE replies with a \packet{tau\_complete} to notify successful completion of the event.

\subsection{Discovered inconsistencies}
\label{subsec:inconsistencies}

Table~\ref{tab:allinc},~\ref{tab:allinc2},~\ref{tab:allinc3},~\ref{tab:allinc4},~\ref{tab:allinc5} and~\ref{tab:allinc6} show a subset of inconsistencies detected by~\system{}. 
\begin{center}

\def\arraystretch{1}
\begin{table*}[ht]
	\centering
    \caption{Subset of inconsistent pairs discovered from 5G.
    ~\boxsec{}: security specification and~\boxnas{}: NAS specification}
	\renewcommand{\arraystretch}{1}
	\fontsize{8}{8}\selectfont
\begin{tabular}{p{0.06\linewidth}p{0.35\linewidth}p{0.35\linewidth}p{0.08\linewidth}p{0.08\linewidth}}
 \hline
 \rule{0pt}{2ex}Pair No. & $\text{Text}_1$ & $\text{Text}_2$ & $\text{Loc}_1$ & $\text{Loc}_2$ \\ [0.5ex] 
 \hline\hline
 \stepcounter{paircount}
 \rule{0pt}{2ex} \thepaircount & The AUSF sends an EAP-Success message to the SEAF \boxcol{together with the SUPI} and the
derived anchor key in the \packet{Nausf\_UEAuthentication\_Authenticate Response}. & The AUSF shall send an EAP Success message to the SEAF inside \packet{Nausf\_UEAuthentication\_Authenticate Response}, which shall forward it transparently to the UE. & \boxsec{B 2.1.1} & \boxsec{6.1.3.1} \\ 
 \hline
 \stepcounter{paircount}
 \rule{0pt}{2ex}\thepaircount & From this time onward the UE shall cipher and integrity protect \boxcol{all NAS signalling messages} with the selected 5GS integrity and ciphering algorithms. & From this time onward, all NAS messages exchanged between the UE and the AMF are sent integrity protected and \boxcol{except for the messages specified in subclause 4.4.5,} all NAS messages exchanged between the UE and the AMF are sent ciphered & \boxnas{5.4.2.3}  & \boxnas{4.4.2.5} \\
 \hline
\stepcounter{paircount}
 \rule{0pt}{2ex}\thepaircount & If this verification is successful, then the UE shall take the received NCC value and save it as stored NCC with the current UE context. The UE shall delete the current AS keys KRRCenc, KUPenc (if available), and KUPint (if available), but keep the current AS key KRRCint key. If the\boxcol{stored NCC value is different from the NCC value} associated with the current KgNB, the UE shall delete the current AS key KgNB. & If the sent \boxcol{NCC value is fresh and belongs to an unused} pair of \{NCC, NH\}, the ng-eNB shall save the pair of \{NCC, NH\} in the current UE AS security context and shall delete the current AS key KgNB. If the sent NCC value is equal to the NCC value associated with the current KgNB, the ng-eNB shall keep the current AS key KgNB and NCC. & \boxsec{6.8.2.1.2} & \boxsec{6.8.2.1.2} \\
 \hline
 \stepcounter{paircount}
 \rule{0pt}{2ex}\thepaircount & Additionally, the UE shall consider the USIM as invalid for the current SNPN until switching off or the UICC containing the USIM is removed In case of SNPN, if the UE is neither registered for onboarding services in SNPN nor performing initial registration for onboarding services in SNPN and the UE supports access to an SNPN using credentials from a credentials holder, the UE shall consider the selected entry of the "list of subscriber data" as invalid until the UE is switched off or the entry is updated.   &  In case of SNPN, if the UE is not performing initial registration for onboarding services in SNPN and the UE supports access to an SNPN using credentials from a credentials holder, the UE shall consider the selected entry of the "list of subscriber data" as invalid for 3GPP access until the UE is switched off, the entry is updated or \boxcol{the timer T3245 expires as described in clause 5.3.19a.2.}   & \boxnas{5.4.1.2.2.11} & \boxnas{5.5.1.2.5} \\
 \hline
 \stepcounter{paircount}
  \rule{0pt}{2ex}\thepaircount & The AUSF sends an EAP-Success message to the SEAF \boxcol{together with the SUPI} and the
derived anchor key in the \packet{Nausf\_UEAuthentication\_Authenticate Response}. & \boxcol{If the authentication was successful},
the KSEAF shall be sent to the SEAF in the \packet{Nausf\_UEAuthentication\_Authenticate Response}. \boxcol{In case the AUSF
received a SUCI} from the SEAF in the authentication request (see sub-clause 6.1.2 of the present document), and
if the authentication was successful, then the AUSF shall also include the SUPI in the
\packet{Nausf\_UEAuthentication\_Authenticate Response} message. & \boxsec{B 2.1.1} & \boxsec{6.1.3.2} \\
 \hline
 \stepcounter{paircount}
 \rule{0pt}{2ex}\thepaircount & Additionally, if EAP based primary authentication and key agreement procedure using EAP-AKA' or 5G AKA based primary authentication and key agreement procedure was performed in the current SNPN, the UE shall consider the USIM as invalid for the current SNPN until switching off or the UICC containing the USIM is removed or the timer T3245 expires as described in clause 5.3.19a.2 If the UE is not registered for onboarding services in SNPN, the UE shall move to 5GMM-DEREGISTERED.NO-SUPI state. & Additionally, if EAP based primary authentication and key agreement procedure using EAP-AKA' or 5G AKA based primary authentication and key agreement procedure was performed in the current SNPN, the UE shall consider the USIM as invalid for the current SNPN until switching off, the UICC containing the USIM is removed or the timer T3245 expires as described in clause 5.3.19a.2 If the UE is not registered for onboarding services in SNPN, the UE shall \boxcol{delete the list} of equivalent PLMNs (if any) and shall enter the state 5GMM-DEREGISTERED.NO-SUPI & \boxnas{5.5.1.3.5} & \boxnas{5.5.2.3.2} \\
 \hline
 \stepcounter{paircount}
 \rule{0pt}{2ex}\thepaircount & If the message was received via 3GPP access and the UE is operating in single-registration mode, the UE shall in addition set the EPS update status to EU3 ROAMING NOT ALLOWED and shall \boxcol{delete any 4G-GUTI,} last visited registered TAI, TAI list and eKSI. & If the message was received via 3GPP access and the UE is operating in single-registration mode, the UE shall in addition set the EPS update status to EU3 ROAMING NOT ALLOWED. &  \boxnas{5.5.1.2.5} & \boxnas{5.5.1.3.5}\\[1ex] 
 \hline
\end{tabular}
\label{tab:allinc}
\end{table*}
\end{center}

\begin{center}

\def\arraystretch{1}
\begin{table*}[ht]
	\centering
    \caption{Subset of inconsistent pairs discovered from 5G (Continued)
    ~\boxsec{}: security specification and~\boxnas{}: NAS specification}
	\renewcommand{\arraystretch}{1}
	\fontsize{8}{8}\selectfont
\begin{tabular}{p{0.06\linewidth}p{0.35\linewidth}p{0.35\linewidth}p{0.08\linewidth}p{0.08\linewidth}}
 \hline
 \rule{0pt}{2ex}Pair No. & $\text{Text}_1$ & $\text{Text}_2$ & $\text{Loc}_1$ & $\text{Loc}_2$ \\ [0.5ex] 
 \hline\hline
 \stepcounter{paircount}
 \rule{0pt}{2ex}\thepaircount & The UE shall delete the list of equivalent PLMNs, \boxcol{reset the registration attempt counter}, store the PLMN identity in the forbidden PLMN list as specified in subclause5.3.13A. For 3GPP access the UE shall enter state 5GMM-DEREGISTERED.PLMN-SEARCH in order to perform a PLMN selection according to 3GPPTS23.122[5], and for non-3GPP access the UE shall enter state 5GMM-DEREGISTERED.LIMITED-SERVICE and perform network selection as defined in 3GPPTS24.502[18]. & The UE shall delete the list of equivalent PLMNs, store the PLMN identity in the forbidden PLMN list as specified in subclause5.3.13A. For 3GPP access the UE shall enter state 5GMM-DEREGISTERED.PLMN-SEARCH in order to perform a PLMN selection according to 3GPPTS23.122[5], and for non-3GPP access the UE shall enter state 5GMM-DEREGISTERED.LIMITED-SERVICE in order to perform network selection as defined in 3GPPTS24.502[18]. & \boxnas{5.5.1.2.5} & \boxnas{5.6.1.5} \\
 \hline

\stepcounter{paircount}
 \rule{0pt}{2ex}\thepaircount & Additionally, the UE shall consider the USIM as invalid for the entry via 3GPP access until switching off or the UICC containing the USIM is removed The UE shall set the counter for "SIM/USIM considered invalid for GPRS services" events and the counter for "SIM/USIM considered invalid for non-GPRS services" events if maintained by the UE, in case of PLMN; or the counter for "the entry for the current SNPN considered invalid for 3GPP access" events in case of SNPN to UE implementation-specific maximum value & \boxcol{If the message has been successfully integrity checked} by the NAS, then the UE shall 1) set the counter for "SIM/USIM considered invalid for GPRS services" events and the counter for "USIM considered invalid for 5GS services over non-3GPP access" events to UE implementation-specific maximum value in case of PLMN if the UE maintains these counters  & \boxnas{5.4.1.2.2.11} & \boxnas{5.5.1.2.5} \\
\hline
\stepcounter{paircount}
 \rule{0pt}{2ex}\thepaircount & Additionally, the UE shall consider the USIM as invalid for the entry via 3GPP access until switching off or the UICC containing the USIM is removed The UE shall set the counter for "SIM/USIM considered invalid for GPRS services" events and the counter for "SIM/USIM considered invalid for non-GPRS services" events if maintained by the UE, in case of PLMN; or the counter for "the entry for the current SNPN considered invalid for 3GPP access" events in case of SNPN to UE implementation-specific maximum value & \boxcol{If the message has been successfully integrity checked} by the NAS, then the UE shall 1) set the counter for "SIM/USIM considered invalid for GPRS services" events and the counter for "USIM considered invalid for 5GS services over non-3GPP access" events to UE implementation-specific maximum value in case of PLMN if the UE maintains these counters & \boxnas{5.4.1.3.5} & \boxnas{5.5.1.3.5} \\
 \hline
 \stepcounter{paircount}
 \rule{0pt}{2ex}\thepaircount & In case of SNPN, if the UE is neither registered for onboarding services in SNPN nor performing initial registration for onboarding services in SNPN and the UE supports access to an SNPN using credentials from a credentials holder, the UE shall consider the selected entry of the "list of subscriber data" as invalid for 3GPP access until the UE is switched off or the entry is updated.  & In case of SNPN, if the UE is not performing initial registration for onboarding services in SNPN and the UE supports access to an SNPN using credentials from a credentials holder, the UE shall consider the selected entry of the "list of subscriber data" as invalid for 3GPP access until the UE is switched off, the entry is updated or \boxcol{the timer T3245 expires} as described in clause5.3.19a.2.  & \boxnas{5.4.1.2.3.1} & \boxnas{5.5.1.2.5} \\
 \hline
 \stepcounter{paircount}
 \rule{0pt}{2ex}\thepaircount & In case of SNPN, if the UE is neither registered for onboarding services in SNPN nor performing initial registration for onboarding services in SNPN and the UE supports access to an SNPN using credentials from a credentials holder, the UE shall consider the selected entry of the "list of subscriber data" as invalid for 3GPP access until the UE is switched off or the entry is updated.  & In case of SNPN, if the UE is not registered for onboarding services in SNPN and the UE supports access to an SNPN using credentials from a credentials holder, the UE shall consider the selected entry of the "list of subscriber data" as invalid for 3GPP access until the UE is switched off, the entry is updated or \boxcol{the timer T3245 expires} as described in clause 5.3.19a.2.     & \boxnas{5.4.1.2.2.11} & \boxnas{5.5.1.3.5} \\
 \hline
 \stepcounter{paircount}
 \rule{0pt}{2ex}\thepaircount & Additionally, if the Updated PEIPS assistance information IE in the CONFIGURATION UPDATE COMMAND message includes a new Paging subgroup ID and the UE is previously assigned a different Paging subgroup ID then, the AMF shall consider \boxcol{both, the old and new Paging subgroup IDs} as valid until the old Paging subgroup ID can be considered as invalid by the AMF & If in the REGISTRATION ACCEPT message a new Paging subgroup ID was assigned to the UE that is different than the old Paging subgroup ID then the network shall initiate the generic UE configuration update procedure; and if no response is received to the paging attempts using the old 5G-S-TMSI from the old 5G-GUTI and the old Paging subgroup ID, the AMF may use the new 5G-S-TMSI from the new 5G-GUTI and the new Paging subgroup ID, if any, for paging, for an implementation dependent number of paging attempts. & \boxnas{5.4.4.6} & \boxnas{5.5.1.3.8} \\
 \hline
 \stepcounter{paircount}
 \rule{0pt}{2ex}\thepaircount & \#13 (Roaming not allowed in this tracking area) The UE shall set the 5GS update status to 5U3 ROAMING NOT ALLOWED (and shall store it according to subclause5.1.3.2.2) and shall \boxcol{delete 5G-GUTI, last visited registered TAI,} \boxcol{TAI list and ngKSI.}   & \#13 (Roaming not allowed in this tracking area) The UE shall set the 5GS update status to 5U3 ROAMING NOT ALLOWED (and shall store it according to subclause5.1.3.2.2) and shall \boxcol{delete the list of equivalent PLMNs} (if available).  & \boxnas{5.5.1.2.5} & \boxnas{5.5.1.3.5} \\
 \hline
\end{tabular}
\label{tab:allinc2}
\end{table*}
\end{center}

\begin{center}

\def\arraystretch{1}
\begin{table*}[ht]
	\centering
    \caption{Subset of inconsistent pairs discovered from 5G (Continued)
    ~\boxsec{}: security specification and~\boxnas{}: NAS specification}
	\renewcommand{\arraystretch}{1}
	\fontsize{8}{8}\selectfont
\begin{tabular}{p{0.06\linewidth}p{0.35\linewidth}p{0.35\linewidth}p{0.08\linewidth}p{0.08\linewidth}}
 \hline
 \rule{0pt}{2ex}Pair No. & $\text{Text}_1$ & $\text{Text}_2$ & $\text{Loc}_1$ & $\text{Loc}_2$ \\ [0.5ex] 
 \hline\hline
 \stepcounter{paircount}
 \rule{0pt}{2ex}\thepaircount & The UE is operating in SNPN access operation mode, the UE shall store the current TAI in the list of "5GS forbidden tracking areas for roaming" for the current SNPN and, if the UE supports access to an SNPN using credentials from a credentials holder, the selected entry of the "list of subscriber data" or the selected PLMN subscription, and enter the state 5GMM-DEREGISTERED.LIMITED-SERVICE or \boxcol{optionally 5GMM-DEREGISTERED.PLMN-SEARCH.}   & the UE is operating in SNPN access operation mode, the UE shall store the current TAI in the list of "5GS forbidden tracking areas for roaming" for the current SNPN and, if the UE supports access to an SNPN using credentials from a credentials holder, the selected entry of the "list of subscriber data" or the selected PLMN subscription, and enter the state 5GMM-DEREGISTERED.LIMITED-SERVICE.   & \boxnas{5.5.1.2.5} & \boxnas{5.5.1.2.5} \\[1ex] 
 \hline
 \stepcounter{paircount}
 \rule{0pt}{2ex}\thepaircount &  \#15 (No suitable cells in tracking area) The UE shall set the 5GS update status to 5U3 ROAMING NOT ALLOWED (and shall store it according to subclause5.1.3.2.2) and shall \boxcol{delete any 5G-GUTI, last} \boxcol{visited registered TAI, TAI list and ngKSI}.  & \#13 (Roaming not allowed in this tracking area) The UE shall set the 5GS update status to 5U3 ROAMING NOT ALLOWED (and shall store it according to subclause5.1.3.2.2).   & \boxnas{5.5.1.3.5} & \boxnas{5.6.1.5} \\

\hline
\stepcounter{paircount}
 \rule{0pt}{2ex}\thepaircount & \#73 (Serving network not authorized). $\cdots$ The UE shall delete the list of equivalent PLMNs, shall \boxcol{reset the registration attempt counter}. For 3GPP access the UE shall enter the state 5GMM-DEREGISTERED.PLMN-SEARCH, and for non-3GPP access the UE shall enter state 5GMM-DEREGISTERED.LIMITED-SERVICE. The UE shall store the PLMN identity in the forbidden PLMN list as specified in subclause 5.3.13A and if the UE is configured to use timer T3245 then the UE shall start timer T3245 and proceed as described in clause 5.3.19a.1. & \#73 (Serving network not authorized). $\cdots$ The UE shall delete the list of equivalent PLMNs and store the PLMN identity in the forbidden PLMN list as specified in subclause 5.3.13A and if the UE is configured to use timer T3245 then the UE shall start timer T3245 and proceed as described in clause 5.3.19a.1 & \boxnas{5.5.1.2.5} & \boxnas{5.6.1.5} \\
 \hline
 \stepcounter{paircount}
 \rule{0pt}{2ex}\thepaircount & \#11 (PLMN not allowed) $\cdots$ The UE shall delete the list of equivalent PLMNs, shall \boxcol{reset the registration attempt counter}. For 3GPP access the UE shall enter the state 5GMM-DEREGISTERED.PLMN-SEARCH, and for non-3GPP access the UE shall enter state 5GMM-DEREGISTERED.LIMITED-SERVICE. The UE shall store the PLMN identity in the forbidden PLMN list as specified in subclause 5.3.13A and if the UE is configured to use timer T3245 then the UE shall start timer T3245 and proceed as described in clause 5.3.19a.1. & \#11 (PLMN not allowed) $\cdots$ The UE shall delete the list of equivalent PLMNs and store the PLMN identity in the forbidden PLMN list as specified in subclause 5.3.13A and if the UE is configured to use timer T3245 then the UE shall start timer T3245 and proceed as described in clause 5.3.19a.1 & \boxnas{5.5.1.2.5} & \boxnas{5.6.1.5}\\
 \hline
 \stepcounter{paircount}
 \rule{0pt}{2ex}\thepaircount & If the UE did not provide a DNN during the PDU session establishment, the UE shall stop the timer T3396 associated with no DNN if it is running & If the UE did not provide a DNN during the PDU session establishment and the request type was \boxcol{different from "initial emergency request"} and \boxcol{different from "existing emergency PDU session"}, the UE shall stop the timer T3396 associated with no DNN if it is running. & \boxnas{6.3.1.2.1} & \boxnas{6.3.3.3} \\
 \hline
 \stepcounter{paircount}
 \rule{0pt}{2ex}\thepaircount & For protection of all RRC messages except \packet{RRCReject} message following the sent \packet{RRCResumeRequest} message, the UE shall derive a $K_{NG-RAN}*$ using the target PCI, target ARFCN-DL/EARFCN-DL and the $K_{gNB}/NH$ based on either a horizontal key derivation or a vertical key derivation as defined in clause 6.9.2.1.1 and Annex A.11/Annex A.12. The UE shall further derive $K_{RRCint}$, $K_{RRCenc}$, $K_{UPenc}$ (optionally), and $K_{UPint}$ (optionally) from the newly derived $K_{NG-RAN}*$. & For protection of all RRC messages except RRC Reject message following the sent RRC Resume Request message, the UE shall derive a $K_{NG-RAN}*$ using the target PCI, target EARFCN-DL and the $K_{gNB}/NH$ based on either a horizontal key derivation or a vertical key derivation as defined in clause 6.9.2.1.1 and Annex A.12. The UE shall further derive $K_{RRCint}$, $K_{RRCenc}$, $K_{UPenc}$ (optionally), and $K_{UPint}$ (optionally) from the newly derived $K_{NG-RAN}*$. \boxcol{Then the UE resets all PDCP COUNTs to 0} and \boxcol{activates the new AS keys in PDCP layer.} & \boxsec{6.8.2.1.3} & \boxsec{6.16.2.3}\\[1ex] 
 \hline
\end{tabular}
\label{tab:allinc3}
\end{table*}
\end{center}

\begin{center}

\def\arraystretch{1}
\begin{table*}[ht]
	\centering
    \caption{Subset of inconsistent pairs discovered from 4G
    ~\boxsec{}: security specification and~\boxnas{}: NAS specification}
	\renewcommand{\arraystretch}{1}
	\fontsize{8}{8}\selectfont
\begin{tabular}{p{0.06\linewidth}p{0.35\linewidth}p{0.35\linewidth}p{0.08\linewidth}p{0.08\linewidth}}
 \hline
 \rule{0pt}{2ex}Pair No. & $\text{Text}_1$ & $\text{Text}_2$ & $\text{Loc}_1$ & $\text{Loc}_2$ \\ [0.5ex] 
 \hline\hline
 \stepcounter{paircountlte}
 \rule{0pt}{2ex}\thepaircountlte & If the algorithms selected by the eNB are different from the algorithms currently used at the target eNB, then the target eNB may \boxcol{take the proper actions} to change to the selected algorithms NOTE: Transferring the ciphering and integrity algorithms used in the source cell to the target eNB in the handover request message is for the target eNB to decipher and integrity verify the RRCReestablishmentComplete message on SRB1 in the potential RRCConnectionRe-establishment procedure. & Even if the AS algorithms used by the source eNB do not match with the target eNB local algorithm priority list the source eNB selected \boxcol{AS algorithms shall take precedence} when running the RRCConnectionRe-establishment procedure. The target eNB and UE should refresh the selected AS algorithms and the AS keys based on local prioritized algorithms after the RRCConnectionRe-establishment procedure NOTE: When the AS algorithms transferred by source eNB are not supported by the target eNB, the target eNB will fail to decipher or integrity verify the RRCReestablishmentComplete message on SRB1. & \boxsec{7.2.4.2.2} & \boxsec{7.4.3} \\ 
 \hline
 \stepcounter{paircountlte}
 \rule{0pt}{2ex}\thepaircountlte &When starting the transition away from EMM-DEREGISTERED state with the intent to eventually transitioning to EMM-REGISTERED state, if no current EPS NAS security context is available in the ME, the ME shall retrieve native EPS NAS security context stored on the USIM if the USIM supports EMM parameters storage and if the stored native EPS NAS security context on the USIM is marked \boxcol{as valid.} & The retrieved native EPS NAS security context with the derived KNASint and KNASenc shall then become the current EPS NAS security context When the ME is transitioning away from EMM-DEREGISTERED state with the intent to eventually transitioning to EMM-REGISTERED state, if the USIM supports EMM parameters storage, the ME shall mark the stored EPS NAS security context on the USIM \boxcol{as invalid.} & \boxsec{7.2.5.2.1} & \boxsec{7.2.5.2.1} \\
 \hline
 \stepcounter{paircountlte}
 \rule{0pt}{2ex}\thepaircountlte &In case of failed integrity check (i.e. faulty or missing MAC-I) is detected after the start of integrity protection, the concerned message shall be discarded. This can happen on the UE side or on the eNB side. & In case of failed integrity check (i.e. faulty or missing NAS-MAC) is detected after the start of NAS integrity protection the concerned message shall be discarded \boxcol{except for some NAS messages} specified in TS 24.301 [9]. For those exceptions the MME shall take the actions specified in TS 24.301 [9] when receiving a NAS message with faulty or missing NAS-MAC. & \boxsec{7.3.2} & \boxsec{8.1.1} \\
 \hline
 \stepcounter{paircountlte}
 \rule{0pt}{2ex}\thepaircountlte & The UE and network shall complete the combined default EPS bearer context activation procedure and the attach procedure \boxcol{before the dedicated} EPS bearer context activation procedure is completed. If EMM-REGISTERED without PDN connection is not supported by the UE or the MME, then the success of the attach procedure is dependent on the success of the default EPS bearer context activation procedure. & The dedicated bearer context activation procedure can be part of the attach procedure or be initiated \boxcol{together with the default} EPS bearer context activation procedure when the UE initiated stand-alone PDN connectivity procedure. If the attach procedure or the default EPS bearer context activation procedure fails, the UE shall consider that the dedicated bearer activation has implicitly failed. & \boxnas{4.2} & \boxnas{6.4.2.1} \\
 \hline
 \stepcounter{paircountlte}
 \rule{0pt}{2ex}\thepaircountlte & From this time onward, all NAS messages exchanged between the UE and the MME are sent integrity protected using the mapped EPS security context, and except for the messages specified in clause 4.4.5, all NAS messages exchanged between the UE and the MME are sent ciphered using the mapped EPS security context During inter-system change from N1 mode to S1 mode in EMM-IDLE mode, if the UE is operating in the single-registration mode and 1) if the tracking area updating procedure is initiated as specified in 3GPP S 24.501 [54], the UE shall transmit a TRACKING AREA UPDATE REQUEST message \boxcol{integrity protected} with the current 5G NAS security context and the UE shall derive a mapped EPS security context (see clause 8.6.1 of 3GPP S 33.501 [56]). & After successful completion of the procedure, all NAS messages exchanged between the UE and the MME are sent integrity protected and except for the messages specified in clause 4.4.5, all NAS messages exchanged between the UE and the MME are sent ciphered 3) If the UE has no current EPS security context and performs a tracking area updating procedure after an inter-system change in idle mode from A/Gb mode to S1 mode or Iu mode to S1 mode, the UE shall send the TRACKING AREA UPDATE REQUEST message \boxcol{without integrity protection and encryption}. & \boxnas{4.4.2.3} & \boxnas{4.4.2.3} \\
 \hline
 \stepcounter{paircountlte}
 \rule{0pt}{2ex}\thepaircountlte & Whenever an ATTACH REJECT message with the EMM cause \#14 "EPS services not allowed in this PLMN" is received by the UE, the chosen PLMN identity shall be stored in the "forbidden PLMNs for GPRS service" and if the UE is configured to use timer T3245 (see 3GPP S 24.368 [15A] or 3GPP S 31.102 [17]) then the UE shall start timer T3245 and proceed as described in clause 5.3.7a. & Additionally, the UE shall delete the list of equivalent PLMNs and reset the attach attempt counter In S1 mode, the UE shall store the PLMN identity in the "forbidden PLMN list" and enter state \boxcol{EMM-DEREGISTERED.PLMN-SEARCH} and if the UE is configured to use timer T3245 (see 3GPP S 24.368 [15] or 3GPP S 31.102 [17]) then the UE shall start timer T3245 and proceed as described in clause 5.3.7a. & \boxnas{5.5.1.1} & \boxnas{5.5.1.3.5} \\
 \hline
 \stepcounter{paircountlte}
 \rule{0pt}{2ex}\thepaircountlte & In addition, the UE shall include Old GUTI type IE with GUTI type set to "native GUTI". If there is no valid GUTI available, the UE shall include the IMSI in the ATTACH REQUEST message; or 2) If the UE supports A/Gb mode or Iu mode or both and i) if the TIN indicates "P-TMSI" and the UE holds a valid native P-TMSI and RAI, the UE shall map the P-TMSI and RAI into the \boxcol{EPS mobile identity IE}, and include Old GUTI type IE with GUTI type set to "mapped GUTI". & In addition, the UE shall include Old GUTI type IE with GUTI type set to "native GUTI"; or 2) if the UE supports A/Gb mode or Iu mode or both, the UE shall handle the Old GUTI IE as follows If the TIN indicates "P-TMSI" and the UE holds a valid native P-TMSI and RAI, the UE shall map the P-TMSI and RAI into \boxcol{the Old GUTI IE}, and include Old GUTI type IE with GUTI type set to "mapped GUTI". & \boxnas{5.5.1.2.2} & \boxnas{5.5.3.2.2} \\
 [1ex] 
 \hline
\end{tabular}
\label{tab:allinc4}
\end{table*}
\end{center}

\begin{center}

\def\arraystretch{1}
\begin{table*}[ht]
	\centering
    \caption{Subset of inconsistent pairs discovered from 4G (Continued)
    ~\boxsec{}: security specification and~\boxnas{}: NAS specification}
	\renewcommand{\arraystretch}{1}
	\fontsize{8}{8}\selectfont
\begin{tabular}{p{0.06\linewidth}p{0.35\linewidth}p{0.35\linewidth}p{0.08\linewidth}p{0.08\linewidth}}
\rule{0pt}{2ex}Pair No. & $\text{Text}_1$ & $\text{Text}_2$ & $\text{Loc}_1$ & $\text{Loc}_2$ \\ [0.5ex] 
 \hline\hline
 \stepcounter{paircountlte}
 \rule{0pt}{2ex}\thepaircountlte & The UE shall consider the USIM as invalid for EPS services and non-EPS services until switching off or the UICC containing the USIM is removed or the timer T3245 expires as described in clause 5.3.7a. Additionally, the UE shall delete the list of equivalent PLMNs and enter state EMM-DEREGISTERED.NO-IMSI. If the message has been \boxcol{successfully integrity checked} by the NAS and the UE maintains a counter for "SIM/USIM considered invalid for GPRS services", then the UE shall set this counter to UE implementation-specific maximum value. & The UE shall consider the USIM as invalid for EPS services until switching off or the UICC containing the USIM is removed or the timer T3245 expires as described in clause 5.3.7a. The UE shall delete the list of equivalent PLMNs and shall enter the state EMM-DEREGISTERED.NO-IMSI. If the UE maintains a counter for "SIM/USIM considered invalid for GPRS services", then the UE shall set this counter to UE implementation-specific maximum value. & \boxnas{5.5.1.2.5} & \boxnas{5.5.2.3.2} \\
 \hline
 \stepcounter{paircountlte}
 \rule{0pt}{2ex}\thepaircountlte & Additionally, the UE shall delete the list of equivalent PLMNs and reset the attach attempt counter In S1 mode, the UE shall store the PLMN identity in the "forbidden PLMN list" and enter state EMM-DEREGISTERED.PLMN-SEARCH and if the UE is configured to use timer T3245 (see 3GPP S 24.368 [15] or 3GPP S 31.102 [17]) then the UE shall start timer T3245 and proceed as described in clause 5.3.7a. & Furthermore, the UE shall \boxcol{delete any GUTI}, last visited registered TAI, TAI \boxcol{ list and KSI}. The UE shall reset the attach attempt counter and shall enter the state EMM-DEREGISTERED.PLMN-SEARCH The UE shall store the PLMN identity in the "forbidden PLMNs for GPRS service" list and if the UE is configured to use timer T3245 (see 3GPP S 24.368 [15A] or 3GPP S 31.102 [17]) then the UE shall start timer T3245 and proceed as described in clause 5.3.7a. & \boxnas{5.5.1.2.5} & \boxnas{5.5.2.3.2} \\
 \hline
 \stepcounter{paircountlte}
 \rule{0pt}{2ex}\thepaircountlte & Additionally, the UE shall reset the registration attempt counter \#12 (Tracking area not allowed) The UE shall set the EPS update status to EU3 ROAMING NOT ALLOWED (and shall store it according to clause 5.1.3.3) and shall \boxcol{delete any GUTI}, last visited registered TAI, TAI list and eKSI. Additionally, the UE shall reset the attach attempt counter In S1 mode, the UE shall store the current TAI in the list of "forbidden tracking areas for regional provision of service" and enter the state EMM-DEREGISTERED.LIMITED-SERVICE. & In addition, the UE shall reset the registration attempt counter \#15 (No suitable cells in tracking area) The UE shall set the EPS update status to EU3 ROAMING NOT ALLOWED (and shall store it according to clause 5.1.3.3). The UE shall reset the tracking area updating attempt counter and shall enter the state EMM-REGISTERED.LIMITED-SERVICE The UE shall store the current TAI in the list of "forbidden tracking areas for roaming". & \boxnas{5.5.1.2.5} & \boxnas{5.5.3.3.5} \\
 \hline
 \stepcounter{paircountlte}
 \rule{0pt}{2ex}\thepaircountlte & Additionally, the UE shall reset the registration attempt counter \#15 (No suitable cells in tracking area) The UE shall set the EPS update status to EU3 ROAMING NOT ALLOWED (and shall store it according to clause 5.1.3.3) and shall delete any GUTI, last visited registered TAI, TAI list and eKSI. Additionally, the UE shall reset the attach attempt counter The UE shall store the current TAI in the list of "forbidden tracking areas for roaming". & Additionally, the UE shall reset the registration attempt counter \#15 (No suitable cells in tracking area) The UE shall set the EPS update status to EU3 ROAMING NOT ALLOWED (and shall store it according to clause 5.1.3.3) and shall delete any GUTI, last visited registered TAI, TAI list and eKSI. Additionally the UE shall reset the attach attempt counter and \boxcol{enter the state} EMM-DEREGISTERED.LIMITED-SERVICE The UE shall store the current TAI in the list of "forbidden tracking areas for roaming". & \boxnas{5.5.1.2.5} & \boxnas{5.5.3.3.5} \\
 \hline
 \stepcounter{paircountlte}
 \rule{0pt}{2ex}\thepaircountlte & It is FFS how to prevent the UE from making repeated attempts at selecting the same satellite access PLMN if there are no other available PLMNs at UE 's location Other values are considered as abnormal cases. The behaviour of the UE in those cases is specified in clause 5.5.1.2.6 . & Additionally, the UE shall reset the registration attempt counter Editor 's note: [IoT\_SAT\_ARCH\_EPS, CR\#3620]. It is FFS how to prevent the UE from making repeated attempts at selecting the same satellite access PLMN if there are no other available PLMNs at UE 's location Other EMM cause values or \boxcol{if no EMM cause IE} is included is considered as abnormal cases. & \boxnas{5.5.1.2.5} & \boxnas{5.5.2.3.2} \\
 \hline
 \stepcounter{paircountlte}
 \rule{0pt}{2ex}\thepaircountlte & Additionally, the UE shall delete the list of equivalent PLMNs and shall enter the state EMM-DEREGISTERED.NO-IMSI. If the message has been \boxcol{successfully integrity checked} by the NAS and the UE maintains a counter for "SIM/USIM considered invalid for GPRS services", then the UE shall set this counter to UE implementation-specific maximum value. & The UE shall consider the USIM as invalid for EPS services until switching off or the UICC containing the USIM is removed or the timer T3245 expires as described in clause 5.3.7a. The UE shall delete the list of equivalent PLMNs and shall enter the state EMM-DEREGISTERED.NO-IMSI. If the UE maintains a counter for "SIM/USIM considered invalid for GPRS services", then the UE shall set this counter to UE implementation-specific maximum value. & \boxnas{5.6.1.5} & \boxnas{5.5.2.3.2} \\ 
 \hline
 \stepcounter{paircountlte}
 \rule{0pt}{2ex}\thepaircountlte & If the UE is operating in single-registration mode, the UE shall in addition set the 5GMM state to 5GMM-DEREGISTERED \#42 (Severe network failure) The UE shall set the EPS update status to EU2 NOT UPDATED, and shall delete any GUTI, last visited registered TAI, TAI list, eKSI, and list of equivalent PLMNs, and set the \boxcol{tracking area update counter} to 5. & If the UE is operating in single-registration mode, the UE shall in addition set the 5GMM state to 5GMM-DEREGISTERED \#42 (Severe network failure) The UE shall set the EPS update status to EU2 NOT UPDATED, and shall delete any GUTI, last visited registered TAI, TAI list, eKSI, and list of equivalent PLMNs. & \boxnas{5.5.3.2.5} & \boxnas{5.5.1.2.5} \\[1ex]
 \hline
\end{tabular}
\label{tab:allinc5}
\end{table*}
\end{center}

\begin{center}

\def\arraystretch{1}
\begin{table*}[ht]
	\centering
    \caption{Subset of inconsistent pairs discovered from 4G (Continued)
    ~\boxsec{}: security specification and~\boxnas{}: NAS specification}
	\renewcommand{\arraystretch}{1}
	\fontsize{8}{8}\selectfont
\begin{tabular}{p{0.06\linewidth}p{0.35\linewidth}p{0.35\linewidth}p{0.08\linewidth}p{0.08\linewidth}}
\rule{0pt}{2ex}Pair No. & $\text{Text}_1$ & $\text{Text}_2$ & $\text{Loc}_1$ & $\text{Loc}_2$ \\ [0.5ex] 
 \hline\hline
 \stepcounter{paircountlte}
 \rule{0pt}{2ex}\thepaircountlte & If the UE finds a suitable GERAN or UTRAN cell, it then proceeds with the appropriate MM and CC specific procedures and the EMM sublayer shall not indicate the abort of the service request procedure to the MM sublayer. Otherwise the EMM sublayer shall indicate the abort of the service request procedure to the MM sublayer NOTE 8: If the UE disables the E-UTRA capability, then subsequent mobile terminating calls could fail If the service request was initiated for CS fallback for emergency call and a CS fallback cancellation request was not received, the UE may attempt to select GERAN or UTRAN radio access technology. & If the UE finds a suitable GERAN or UTRAN cell, it then proceeds with the appropriate MM and CC specific procedures and the EMM sublayer shall not indicate the abort of the service request procedure to the MM sublayer. Otherwise the EMM sublayer shall indicate the abort of the service request procedure to the MM sublayer, and the UE shall also set the EPS update status to EU2 NOT UPDATED and enter the state \boxcol{EMM-REGISTERED.ATTEMPTING-TO-UPDATE} & \boxnas{5.6.1.5} & \boxnas{5.6.1.5} \\
 \hline
\stepcounter{paircountlte}
\rule{0pt}{2ex}\thepaircountlte & If the UE finds a suitable GERAN or UTRAN cell, it then proceeds with the appropriate MM and CC specific procedures and the EMM sublayer shall not indicate the abort of the service request procedure to the MM sublayer. Otherwise the EMM sublayer shall indicate the abort of the service request procedure to the MM sublayer NOTE 8: If the UE disables the E-UTRA capability, then subsequent mobile terminating calls could fail If the service request was initiated for CS fallback for emergency call and a CS fallback cancellation request was not received, the UE may attempt to select GERAN or UTRAN radio access technology. & If the UE finds a suitable GERAN or UTRAN cell, it then proceeds with the appropriate MM and CC specific procedures and the EMM sublayer shall not indicate the abort of the service request procedure to the MM sublayer. Otherwise the EMM sublayer shall indicate the abort of the service request procedure to the MM sublayer, and the UE shall also set the EPS update status to EU2 NOT UPDATED and enter the state \boxcol{EMM-REGISTERED.ATTEMPTING-TO-UPDATE} & \boxnas{5.6.1.5} & \boxnas{5.6.1.5} \\
 \hline
 \stepcounter{paircountlte}
 \rule{0pt}{2ex}\thepaircountlte & The following abnormal cases can be identified a) Expiry of timer T3485 On the first expiry of the timer T3485, the MME shall resend the ACTIVATE DEFAULT EPS BEARER CONTEXT REQUEST and shall reset and restart timer T3485. This retransmission is repeated four times, i.e. on the fifth expiry of timer T3485, the MME shall release possibly allocated resources for this activation and shall abort the procedure & The following abnormal cases can be identified a) Expiry of timer T3485 On the first expiry of the timer T3485, the MME shall resend the ACTIVATE DEDICATED EPS BEARER CONTEXT REQUEST and shall reset and restart timer T3485. This retransmission is repeated four times, i.e. on the fifth expiry of timer T3485, the MME shall abort the procedure, release any resources allocated for this activation and \boxcol{enter the state BEARER CONTEXT INACTIVE} & \boxnas{6.4.1.6} & \boxnas{6.4.2.6} \\
 \hline
 \stepcounter{paircountlte}
 \rule{0pt}{2ex}\thepaircountlte & When the UE is switched off, when the USIM is removed, or \boxcol{when there is a change in the value} indicated by the network in \boxcol{the 15 bearers bit} of the EPS network feature support IE, the UE shall clear all previous determinations representing PLMNs maximum number of EPS bearer contexts in S1 mode. & When the UE is switched off or when the USIM is removed, the UE shall clear all previous determinations representing any PLMN 's maximum number of EPS bearer contexts in S1 mode. & \boxnas{6.5.1.4.3} & \boxnas{6.5.3.4.2} \\
 \hline
 \stepcounter{paircountlte}
 \rule{0pt}{2ex}\thepaircountlte & Tracking area updating and detach procedure collision EPS detach containing detach type "re-attach required" or "re-attach not required": If the UE receives a DETACH REQUEST message before the tracking area updating procedure has been completed, the tracking area updating procedure shall be aborted and the detach procedure shall be progressed. & Change of cell into a new tracking area If a cell change into a new tracking area that is not in the stored TAI list occurs before the UE initiated detach procedure is completed, the UE proceeds as follows: 1) If the detach procedure was initiated for reasons other than removal of the USIM or the UE is to be switched off, the detach procedure shall be aborted and re-initiated after successfully performing a tracking area updating procedure & \boxnas{5.5.3.2.6} & \boxnas{5.5.2.2.4} \\
 \hline
 \stepcounter{paircountlte}
 \rule{0pt}{2ex}\thepaircountlte & The UE shall store the current TAI in the list of "forbidden tracking areas for roaming". If the ATTACH REJECT message is not integrity protected, the UE shall memorize the current TAI was stored in the list of "forbidden tracking areas for roaming" for non-integrity protected NAS reject message. Additionally, the UE shall enter \boxcol{the state EMM-DEREGISTERED.LIMITED-SERVICE} & The UE shall store the current TAI in the list of "forbidden tracking areas for roaming". If the ATTACH REJECT message is not integrity protected, the UE shall memorize the current TAI was stored in the list of "forbidden tracking areas for roaming" for non-integrity protected NAS reject message. & \boxnas{5.5.1.2.5} & \boxnas{5.5.1.3.5} \\
 \hline
 \stepcounter{paircountlte}
 \rule{0pt}{2ex}\thepaircountlte & From this time onward the UE shall cipher and integrity protect all NAS signalling messages with the selected NAS
ciphering and NAS integrity algorithms.  & From this time onward, all NAS messages exchanged between the UE and the
MME are sent integrity protected and \boxcol{except for the messages} specified in clause 4.4.5, all NAS messages
exchanged between the UE and the MME are sent ciphered & \boxnas{5.4.3.3} & \boxnas{4.4.2.3} \\
 \hline
 \stepcounter{paircountlte}
 \rule{0pt}{2ex}\thepaircountlte & If the UE finds a suitable GERAN or UTRAN cell, it then proceeds with the appropriate MM and CC specific procedures; otherwise the EMM sublayer shall indicate the abort of the EMM procedure to the MM sublayer . & If the UE finds a suitable GERAN or UTRAN cell, it then proceeds with the appropriate MM and CC specific procedures; otherwise if there is a \boxcol{CS fallback emergency call} or \boxcol{CS fallback call pending}, the EMM sublayer shall indicate the abort of the EMM procedure to the MM sublayer . & \boxnas{5.5.1.3.6} & \boxnas{5.5.3.3.4.3} \\[1ex]
\hline
\end{tabular}
\label{tab:allinc6}
\end{table*}
\end{center}